\documentclass[preprint,showpacs,preprintnumbers,amsmath,amssymb]{revtex4-2}

\usepackage{graphicx}\graphicspath{./graph/}%
\usepackage{dcolumn}
\usepackage{bm}
\usepackage{color}
\usepackage{verbatim}
\usepackage{amssymb}
\usepackage{amsfonts}
\usepackage{latexsym}
\usepackage{epstopdf}
\usepackage{url}
\usepackage{float}
\usepackage{placeins}
\definecolor{red}{rgb}{1,0,0}
\definecolor{green}{rgb}{0,1,0}
\definecolor{blue}{rgb}{0,0,1}

\newcommand{\be}{\begin{equation}}
\newcommand{\ee}{\end{equation}}
\newcommand{\bea}{\begin{eqnarray}}
\newcommand{\eea}{\end{eqnarray}}
\newcommand{\bdm}{\begin{displaymath}}
\newcommand{\edm}{\end{displaymath}}

\newcommand\ra{\rightarrow}

\begin{document}
\title{Equilibrium  of an Arbitrary Bunch Train \\with Cavity Resonators and Short Range Wake: \\Enhanced Iterative
Solution with Anderson Acceleration}
\author{Robert Warnock }
\email{warnock@slac.stanford.edu}
\affiliation{SLAC National Accelerator Laboratory, Stanford University, Menlo Park, CA 94025, USA}
\affiliation{Department of Mathematics and Statistics, University of New Mexico, Albuquerque, NM 87131, USA}

\begin{abstract}
This paper continues the work of two previous treatments of bunch lengthening by a passive harmonic cavity in an electron
storage ring. Such cavities, intended to reduce the effect of Touschek scattering, are a feature of  fourth generation synchrotron light sources. The charge densities in the equilibrium state are given by solutions of coupled Ha\"issinski equations, which are nonlinear integral equations. If the only wake fields are from cavity resonators, the unknowns can be the Fourier transforms of bunch densities at the resonator
frequencies. The solution scheme based on this choice of unknowns proved to be deficient at the design current when multiple resonators were included. Here we return to the conventional formulation of Ha\"issinski equations in coordinate space, the unknowns being charge densities at mesh points on a fine grid.  This system would be awkward to solve by the Newton method used previously, because the Jacobian matrix is very large. Here a new solution is described, which is both Jacobian-free and much simpler. It is based on an elementary fixed point iteration, accelerated by Anderson's method. The  scheme is notably fast and  robust, accommodating even the case of extreme over-stretching at  current
far beyond the design value.  The Anderson method is promising for many problems in accelerator theory and beyond, since it is quite simple and can be used to attack all kinds of nonlinear and linear integral and differential equations. Results are presented for ALS-U, with updated design parameters. The model includes harmonic and main r.f. cavities, compensation of beam loading of the main cavity by adjustment of the generator voltage, and a realistic short range wake field (rather than the broad-band resonator wake invoked previously).
\end{abstract}
\maketitle
\section{Introduction}
The problem is to determine the longitudinal charge distributions of an arbitrary bunch train in an electron storage ring, in the state of equilibrium. This equilibrium could exist theoretically, but be unstable in practice. Any coupling to transverse degrees of freedom is ignored.
The train is arbitrary in the sense that there can be any distribution of gaps (unfilled buckets), and any distribution of bunch current along the train.

Assuming that the collective motion is governed by coupled Vlasov-Fokker-Planck equations, the equilibrium state is determined by coupled Ha\"issinski equations \cite{prabI}. The coupling arises from the long range wake fields of high-Q cavity resonators. Every bunch in the train contributes to the excitation of these wakes, and thereby influences all other bunches and even itself (by a very small amount).  There is also the short range wake field from geometric aberrations in the vacuum chamber, affecting only the bunch that excites it. An achievement of the present work is to include this effect accurately, which has not been done before in a multi-bunch framework.

In \cite{prabI} the problem was solved for the simplest model, in which the only wake field comes from a single passive higher harmonic cavity (HHC). In \cite{prabII} the model was extended to include the wake field (beam loading) of the main accelerating cavity (MC), and also the short range wake field (SR),
roughly approximated by a broad band resonator model.  The r.f. generator voltage was adjusted by a least-squares algorithm so that the sum
of the generator voltage and the induced voltage of the main cavity closely approximated the desired accelerating voltage, in amplitude and phase.
Contrary to the supposition of \cite{prabI}, the main cavity compensated in this way played a substantial role, spoiling to some extent the desired effect of HHC. Also, the consequence of SR was not negligible.

Both \cite{prabI} and \cite{prabII} were based on a formulation of the Ha\"issinski
equations in which the unknowns are the Fourier transforms of the bunch densities at the frequencies of the cavity resonators.
(Of course this is only possible if the  entire wake field is due to resonators, narrow- or broad-band.) Then the number of real unknowns
is $n_u=2n_bn_r$, where $n_b$ is the number of bunches and $n_r$ the number of resonators. For ALS-U we have $n_b=284$ and $n_r=3$ for the model with HHC+MC+SR, thus $n_u= 1704$. Newton's method is readily feasible for solution of a system of this size or even much larger. In fact the method worked beautifully in \cite{prabI}, where $n_u=656$, but failed to converge for the full range of parameters desired in \cite{prabII},
with $n_u=1704$ or larger.

A possible way to avoid the divergence might be to return to the conventional formulation of the Ha\"issinski equations in coordinate space ($z$-space). The single-bunch equation in $z$-space, discretized on a mesh, is solved very robustly by Newton's method, even at currents far beyond realistic values \cite{bobkarl}. Hoping for a similar success in the multi-bunch case, one encounters the problem of a very large Jacobian matrix. With
 a mesh of 100 cells and 284 bunches the dimension of the matrix is 28684 $\times$ 28684, which is uncomfortable if not impossible on a standard PC. Moreover, other light source designs have more than 1000 bunches. Instead of a full Newton method, one could consider more economical quasi-Newton
procedures such as Broyden's method \cite{broyden}, \cite{kelley}.

Fortunately, a very simple and effective $z$-space solution turned up in the guise of a relaxed fixed point iteration, suggested by He, Li, Bai, and Wang \cite{hefei}. This is Jacobian-free, and involves little  calculation beyond repeated evaluations of the potential function that appears in the exponent of the Ha\"ssinski operator. This was successfully applied with parameters for ALS-U and other rings in \cite{hefei}, and I have
verified the success for ALS-U.
Many of the problems posed in \cite{prabII} were solved in a simpler way by this method, but there were still some failures of convergence in cases of interest. Convergence of the method is slow, the number of iterations required being of order 100, but the total computation time is nevertheless modest.

This development turned my attention away from Newton-type methods and toward Jacobian-free fixed point iterations. There is a long history of efforts to accelerate iterative sequences \cite{brezinski}. One of particular interest is Anderson's proposal of 1965 \cite{anderson},
\cite{walker-ni}, \cite{fang-saad}. It has the potential both to cure divergence and to promote fast convergence. Remarkably, it does both in our problem, providing a very fast and robust solution throughout the parameter domain of interest.

Section \ref{section:rfp} describes  the  relaxed fixed point iteration. The discussion leads naturally to the continuation
method, which is a more standard approach to nonlinear equations and a technique that can be related to Anderson acceleration. Section \ref{section:aa} introduces Anderson acceleration. Section \ref{section:results}
presents results for ALS-U, with parameters from the latest design report, somewhat different from those of our previous papers.
Section \ref{section:relation} treats the relation of Anderson acceleration to Broyden's quasi-Newton method. Section \ref{section:fine}
presents conclusions and the outlook for future work.
\section{ Relaxed fixed point iteration and the continuation method \label{section:rfp}}
We wish to solve $n$ equations in $n$ unknowns, written compactly as
\be
x=g(x)\ ,\quad g: R^n\ra R^n\ ,                  \label{gdef}
\ee
where $g$ may be linear but is nonlinear in general. To relate to later discussions we suppose that $g$ has  continuous first derivatives, as is true in our examples,
although no derivatives appear in the  numerical work. The function $g$ will be called the {\it basic map}. The elementary fixed point iteration, or method of
successive substitutions, tries to construct a solution by starting with some guess $x_0$ and forming a sequence $\{ x_k\}$ as
\be
x_{k+1}=g(x_k)\ , \quad k=0,1,\cdots\ ,                \label{efp}
\ee
hoping that the sequence will converge to a solution $x$.
If $g$ maps a ball $B_r=\{x|~\| x\|< r~\}$ into itself, and  reduces  the distance between any two points in the ball,
\be
  \|g(x_1)-g(x_2\|< \beta\|x_1-x_2\|\ ,\quad {\rm all}~~ x_1, x_2 \in B_r\ ,\quad 0<\beta<1\ , \label{contractive}
\ee
then the Contraction Mapping Theorem ensures that the sequence converges to a solution, the only solution in $B_r$, for any $x_0\in B_r$.
Here $\|\cdot\|$ can be any norm, but for analytic estimates of $\beta$ a convenient choice is the maximum absolute value of components of $x$:
$\|x\|=\max_i |x^i|$.

The discretized Ha\"ssinski system has the contractive property for sufficiently small beam current, so we have the assurance of a unique solution
at low current. Numerical calculations show that the sequence diverges for the larger currents of interest, so we seek a better
algorithm.

We look for a better map $h(x)$, which ought to  generate a sequence that will converge, at least for some appropriate $x_0$. The {\it relaxed}
or {\it damped} fixed point iteration makes $h(x)$ from $g(x)$ in the simplest imaginable way,
\be
h(x)=\alpha g(x)+(1-\alpha)x\ , \qquad 0 < \alpha <1\ .      \label{relaxed}
\ee
That is, if $g$ produces too much change in $x$, reduce its contribution and use the current $x$ itself for the rest of the iterate. It could be said  that damping rather than relaxation  is more descriptive of the process.

  He {\it et al.} in \cite{hefei} adopted this procedure to solve the coupled Ha\"issinski equations in the $z$-space formulation, and found it to be remarkably effective. They called it a {\it relaxation iteration}. The damping parameter $\alpha$ was chosen by experiment. At high values of current a relatively small value is required
 for convergence, say $\alpha=0.1$.

 It looks as though He {\it et al.} generalized freely from the linear case, since their bibliography on the source of the method refers only
  to that case. When $g$ is linear relaxation is widely used in the SOR algorithm, Successive Over-Relaxation, a variant of the Gauss-Seidel method intended to accelerate convergence.
 I have not seen much notice of the nonlinear application in the literature of numerical analysis, although one can find it in the context
  of particular problems. See for instance Section III of \cite{dederichs} where it is called ``simple mixing".

A bit more insight into (\ref{relaxed}) accrues if we invoke a differential equation. Define $f(x)=g(x)-x$ and consider the system of ordinary
differential equations,
\be
\frac{dx}{dt}=f(x)\ ,\qquad  x(0)=x_0\ .                      \label{ode}
\ee
One might hope to find a solution of $f(x)=0$ as the limit of an asymptotically constant trajectory $x(t)$ as $t\ra \infty$. Euler's
method applied to (\ref{ode}) gives a sequence $\{ x_k\}$ defined by
\be
\frac{x_{k+1}-x_k}{\Delta t}=f(x_k)\ ,\qquad x_{k+1}=\Delta t ~g(x_k)+(1-\Delta t)~x_k\ .   \label{euler}
\ee
Thus an attempt to find a constant asymptote by Euler's method is the same as trying to find a solution by the relaxed fixed point iteration,
with $\alpha=\Delta t$.

There is no proof that a solution of $f(x)=0$ is really to be found as the constant asymptote of a solution of (\ref{ode}).  A more certain
relation to a differential equation, explored in the literature, is obtained by considering a homotopy connecting an equation with known solution $x_0$
to the equation of interest \cite{ortega}, \S 7.5. Suppose that $H: R^n\times R^1\ra R^n$ is a smooth function of both variables such that
\be
H(x_0,t_0)=0\ ,\qquad H(x,t_1)=f(x)\ .  \label{initfinal}
\ee
Then consider the trajectory $x(t)$ defined by $H(x(t),t)=0$. Differentiating we find
\be
H_x(x,t)\frac{dx}{dt}+H_t(x,t)=0\ . \label{diff}
\ee
As long as the inverse of the Jacobian $H_x$ exists, we have the differential equation in explicit form,
\be
\frac{dx}{dt}= -H_x^{-1}(x,t)H(x,t)\ , \quad x(t_0)=x_0\ .     \label{homode}
\ee
If this equation has a solution extending from $t_0$ to $t_1$, we have achieved a
solution of $f(x)=0$ as $x=x(t_1)$,  according to (\ref{initfinal}). A procedure along these lines is called
a {\it continuation method}.

There are of course myriad ways to choose $H$. An especially useful choice is
\bea
&&H(x,t)=f(x)-e^{-t}f(x_0)\ ,\qquad t_0=0\ ,~~ t_1=\infty\ , \nonumber\\
&&H_x(x,t)=f_x(x)\ ,\quad H_t(x,t)=e^{-t}f(x_0)\ .       \label{hchoice}
\eea
Since on the trajectory $e^{-t}f(x_0)=f(x)$, the equation (\ref{homode}) becomes
\be
\frac{dx}{dt}=-f_x^{-1}(x)f(x)\ .   \label{newtde}
\ee

When Euler's method is applied to (\ref{newtde}) we get
\be
x_{k+1}=x_k-\Delta t~f_x^{-1}(x_k)f(x_k)\ ,                    \label{dampednewt}
\ee
which is the damped Newton method with damping factor $\Delta t$. It becomes the full Newton method for $\Delta t=1$.
If $F(x)=-f_x(x)^{-1}f(x)$ obeys a Lipschitz condition the differential equation (\ref{newtde}) is subject to standard existence
theorems. Boggs \cite{boggs} has explored the use of more sophisticated integrators of (\ref{newtde}), with the goal of approaching thr asymptote
more quickly.

Broyden's method gives a way to  approximate $f_x^{-1}(x_k)$, starting with a value for $f_x^{-1}(x_0)$ \cite{kelley}. The update from step $k$ to step $k+1$ is obtained  by adding a rank-1 matrix. With the definition $G_k\approx f_x^{-1}(x_k)$ the update takes the form
\be
G_{k+1}=G_k+(\Delta x_k-G_k\Delta f_k)\frac{\Delta f_k^T}{\Delta f_k^T\Delta f_k}\ ,\quad \Delta v_k=v_{k+1}-v_k\ ,  \label{broyden}
\ee
where the row vector $v^T$ is the transpose of a column vector $v$. This is called Broyden's second method. His first method approximates the Jacobian itself in a similar way.

Sometimes it is adequate to take $f_x(x_0)=-I$ where $I$ is the unit matrix, which is equivalent to using the relaxed iteration of (\ref{euler}) for the first step. With this reasonable choice and (\ref{broyden}) one can carry out an approximate version of the Newton iteration
(\ref{dampednewt}), often to good effect. The undamped iteration would be preferred, but damping could be needed for convergence.

\section{Anderson Acceleration \label{section:aa}}
Again we wish to solve $x=g(x)$.
Step $k$ of Anderson's iteration  makes use of the current and previous   evaluations of the map, $g(x_j)\ ,\ j=k,k-1,\cdots  $.
These evaluations contain valuable information. The update $x_{k+1}$ is formed from a favorable linear combination of the $g(x_j)$.

With a given start $x_0$ we employ the following notations for $k=0,1,\cdots$:
\be
 g_k=g(x_k)\ ,\quad f_k=g_k-x_k\ ,\quad \|u\|^2=\sum_{i=1}^n (u^i)^2\ ,\quad u=(u^1,\cdots,u^n)\ . \label{somedefs}
\ee
Also choose an integer $m\ge 1$ and define
\be
m_k=\min(k,m)\ ,  \label{mk}
\ee
which will be the number of previous map evaluations used at the $k$-th step, not more than $m$.

To find a good linear combination of the $g_j$, Anderson finds the coefficients in a minimal linear combination of the $f_j$. That is, he
solves the constrained linear least-squares problem
\be
(\alpha_0^k,\alpha_1^k,\cdots,\alpha_{m_k}^k)=\arg\min \|~\sum_{j=0}^{m_k}\alpha_j^kf_{k-m_k+j}~\|^2\ ,\qquad \sum_{j=0}^{m_k}\alpha_j^k=1\ .   \label{lsq}
\ee
Then the next iterate is taken to be
\be
x_{k+1}=\sum_{j=0}^{m_k}\alpha_j^kg_{k-m_k+j}\ .  \label{update}
\ee
For $k=0$ the constraint alone determines the minimum, so that $\alpha_0^0=1$ and $x_1=g(x_0)$.

Anderson allowed extra flexibility by introducing a relaxation parameter $\beta_k$, with a corresponding update
\be
x_{k+1}=\beta_k\sum_{j=0}^{m_k}\alpha_j^kg_{k-m_k+j}+(1-\beta_k)\sum_{j=0}^{m_k}\alpha_j^kx_{k-m_k+j}\ .  \label{relaxedupdate}
\ee
In this scheme $x_1=\beta_0g(x_0)+(1-\beta_0)x_0$, which is to say that the iteration starts with simple mixing. In view of the
partial success of simple mixing, this would seem to be a good choice, at least for the first step. At later iterations one might put $\beta_k=1$.

It is convenient, both for the calculation and for some steps in analysis, to recast the minimization problem without constraints.
That is accomplished merely by a linear change of variables; see \cite{walker-ni}, Eq.(3.1)ff. Define new constants $\gamma_i^k$ such that
\be
\alpha_0^k=\gamma_0^k\ ,\qquad \alpha_j^k=\gamma_j^k-\gamma_{j-1}^k\ ,\quad 1\le j\le m_k-1\ ,\qquad \alpha_{m_k}^k=1-\gamma_{m_k-1}^k\ . \label{gamma}
\ee
Now the sum of the $\alpha_j^k$ is 1 for any choice of the $\gamma_j^k$, and the unconstrained minimization takes the form
\be
(\gamma_0^k,\gamma_1^k,\cdots,\gamma_{m_k}^k)=\arg\min \|~f_k+\sum_{j=0}^{m_k-1}\gamma_j^k(f_{k-m_k+j}-f_{k-m_k+j+1})~\|^2\ .   \label{unconstr}
\ee
Correspondingly, the next iterate is
\be
x_{k+1}=g_k+\sum_{j=0}^{m_k-1}\gamma_j^k(g_{k-m_k+j}-g_{k-m_k+j+1})\ .   \label{gupdate}
\ee
Of course, with relaxation this becomes
\be
x_{k+1}=\beta_k\big(g_k+\sum_{j=0}^{m_k-1}\gamma_j^k(g_{k-m_k+j}-g_{k-m_k+j+1})\big)+
(1-\beta_k)\big(x_k+\sum_{j=0}^{m_k-1}\gamma_j^k(x_{k-m_k+j}-x_{k-m_k+j+1})\big)\ .
\ee

The relation of Anderson's method to Broyden's algorithm is discussed in Section \ref{section:relation}.

\section{Results with ALS-U parameters\label{section:results}}
Parameters considered in the preliminary design report for ALS-U \cite{pdr} of October 2020 are listed in Table 1.
\begin{table}
\begin{center}
\caption{~~Parameters from preliminary design report for ALS-U}
\vskip .2cm
\setlength{\tabcolsep}{.25in}
\begin{tabular} {l|c|c}
Ring circumference& $C$&196.5 ~m\\
Beam energy&$E_0$&2~ GeV\\
Average bunch current&$I_{\rm avg}$&500~mA\\
Momentum compaction&$\alpha$&$2.025\times10^{-4}$\\
Natural energy spread&$\sigma_\delta$&$1.02\times10^{-3}$\\
Natural rms bunch length&$\sigma_{z0}$&3.9 mm\\
Energy loss per turn (with ID's)&$U_0$&315 - 330 keV\\
Harmonic number&$h$&328\\
Main cavity frequency&$f_1$&500.390 MHz\\
Main cavity voltage&$V_1$&600 kV\\
Harmonic cavity harmonic number &$3$&\\
Harmonic cavity shunt impedance&$R_s$ (high $R/Q$)&1.9 M$\Omega$\\
Harmonic cavity  quality factor&$Q$ (high $R/Q$)&$2.4\times10^4$\\
Harmonic cavity detuning&$f_r-3f_1$ (high $R/Q$)&317.80 kHz\\
Harmonic cavity shunt impedance&$R_s$ (low $R/Q$)&1.4 M$\Omega$\\
Harmonic cavity  quality factor&$Q$ (low $R/Q$)&$3.4\times10^4$\\
Harmonic cavity detuning&$f_r-3f_1$ (low $R/Q$)&164.74 kHz\\
Main cavity shunt impedance (sum of 2)&$R_s$ (unloaded)& 9.8 M$\Omega$\\
Main cavity quality factor&$Q$ (unloaded)& $3.6\times 10^4$\\
Main cavity detuning&$f_r-f_1$&-94.729 kHz\\
Main cavity coupling parameter& $\beta$ (optimum) & 9.983\\
Main cavity coupling parameter& $\beta$ (ALS heritage) & 3.1\\
\end{tabular}
\end{center}
\label{table: table1}
\end{table}
These parameters differ considerably from those adopted  in references \cite{prabI}, \cite{prabII}, and \cite{hefei}, so part of the motivation for this report is to bring the study up to date. Two choices for the harmonic cavity parameters are contemplated, called the high $R/Q$ and low $R/Q$ options, which are alleged to have different implications for stability issues. The stability is of course important, but does not concern us here.

At last notice the coupling coefficient $\beta$ for the main rf cavity was still an undetermined feature of the design. The present ALS cavities might be used if their coupling
could be increased enough to control the dc Robinson instability.  For consistency with the calculations of \cite{pdr} we take the ``optimum" value of Table 1, ~$\beta= 9.983$.   Appropriate values of impedance and quality factor for the calculation are the loaded values, $R_{sL}=R_s/(1+\beta)\ ,\ Q_L=Q/(1+\beta)$.  We use the main cavity detuning from the table, which  realizes the ``compensated condition"  given by Eqs. (3.79) and (3.80) in \cite{pdr}.

The table in \cite{pdr} gives $U_0=330$ keV with insertion devices, but the reported calculations to be compared to ours have $U_0=315$ keV, so we choose the latter.

The object is to solve the discretized coupled Ha\"issinski system, written compactly as
\be
f(\rho)=0\ , \quad f: R^n\rightarrow R^n \ ,  \label{abhais}
\ee
 Supposing that the mesh for discretization of each bunch density has $n_m$ points, the vector $\rho$ with $n=n_mn_b$ components consists of $n_b$ densities evaluated at the mesh points:
\be
\rho=
\big[\rho_1(z_1),\rho_1(z_2),\cdots,\rho_1(z_{n_m}),\cdots,\rho_{n_b}(z_1),\rho_{n_b}(z_2),\cdots,\rho_{n_b}(z_{n_m})\big]\ .  \label{rhovec}
\ee
Similarly,
\be
f=\big[f_1(z_1),f_1(z_2),\cdots,f_1(z_{n_m}),\cdots,f_{n_b}(z_1),f_{n_b}(z_2),\cdots,f_{n_b}(z_{n_m})\big]\ , \label{fvec}
\ee
where
\be
f_i(z_j)=\frac{1}{A_i}\exp(-\mu U_i(z_j,\rho)-\rho_i(z_j)\ .    \label{fdef}
\ee
The denominator $A_i$ is a normalization factor, the discretized integral of the numerator. In the exponent
$U_i(z_j,\rho)$ is the potential well seen by the $i$-th bunch, and $\mu$ is the constant of Eq.(49) in \cite{prabI}.
In the following $U_i$ consists of the expression defined in Eqs. (51), (57), and (60) of \cite{prabI}, plus the integral of the
short range wake potential convolved with $\rho_i$, as follows:
\be
U_{i}^{sr}(z ) = \frac{1}{f_1}\int_{-\Sigma}^\Sigma S(z-\zeta)\rho_i(\zeta)d\zeta\ ,\qquad S(z)=\int_{\zeta_0}^zW(\zeta)d\zeta\ . \label{usr}
\ee
The wake potential $W(z)$ , from detailed modeling by Dan Wang, is plotted in Fig.13 of \cite{prabII}. It is zero for $z<\zeta_0$,
where $\zeta_0$ is a small fraction of the bunch length, arising from the non-zero length of the drive bunch in the wake field simulation.

The solution vector $x$ of the previous section is identified with $\rho$ and the map vector $g$ is  from the first term in (\ref{fdef});
that is $g=f+\rho$.

The least-squares step in the Anderson algorithm is done with the code {\bf dgels } from the Intel Math Kernel Library. This solves the normal equation using the QR decomposition, which is recommended in \cite{walker-ni}. We take $m_k=\min(k,\infty)=k$ so that at the $k$-th iterate the current evaluation and
all previous evaluations of the map $g$ are employed. About the same results are obtained with a sufficiently large limit on the number
used, say with $m_k=\min(k,8)$, but this is bothersome to verify and gives no appreciable saving in computation time. The time for the least-squares step is negligible.

All results and CPU times are for a serial code in Fortran, running on a laptop. The code is arranged so that the result of any run
can be taken as an initial guess for the next run. A result for a complete fill can then be used to initiate a run with a partial fill,
or with a smaller detuning, or with a new wake component included, or with  smaller error tolerances, etc.

In contrast to the algorithms used in \cite{prabI} and \cite{prabII},  no continuation in current from small initial values is needed
to achieve convergence. In spite of strong nonlinearities convergence is found immediately at the design current and even at much higher values.
\begin{figure}[htb]
   \centering
   \includegraphics[width=0.6\linewidth]{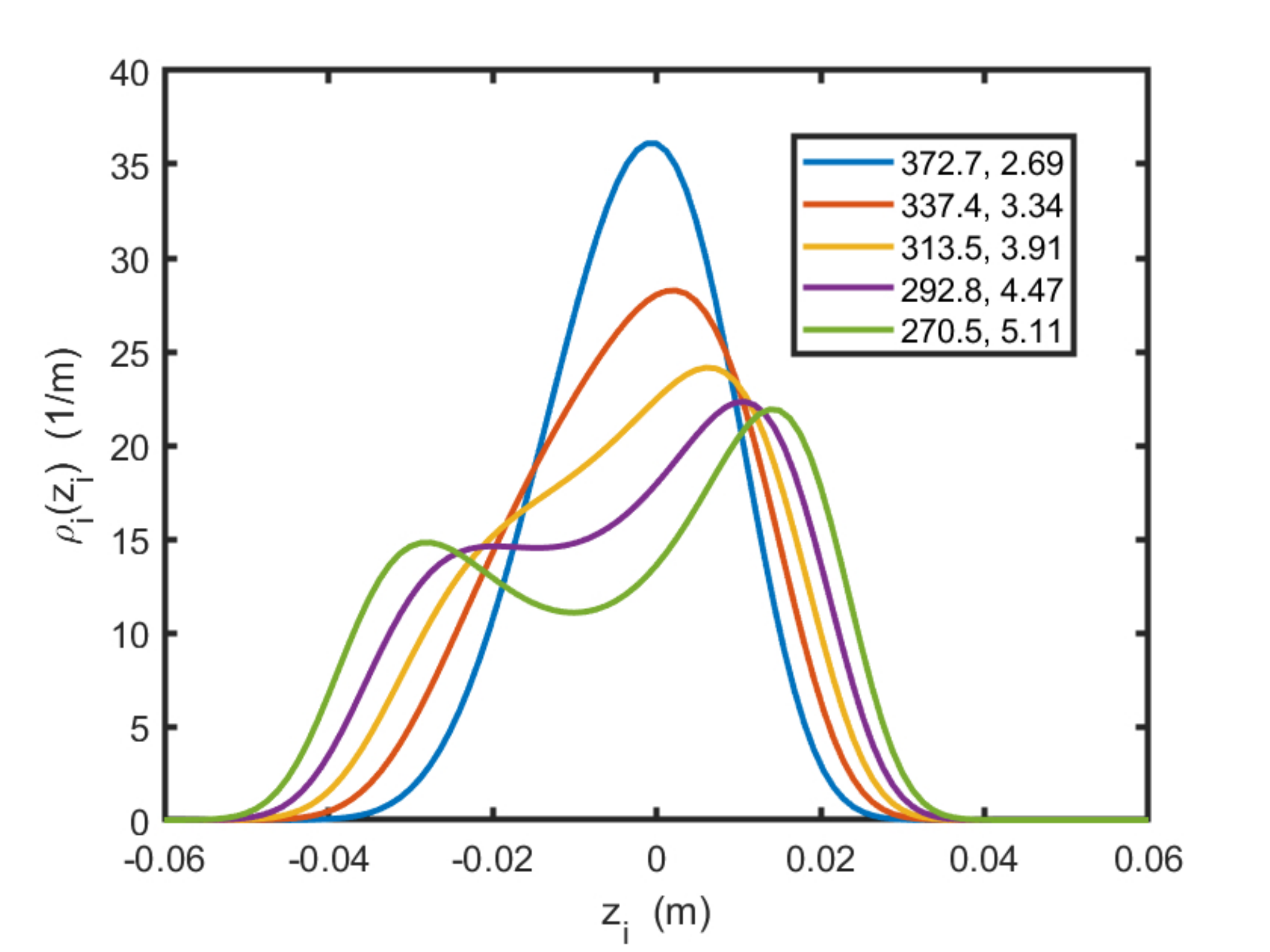}
  \caption{Complete fill, HHC only, high $R/Q$. The legend gives the detuning $\delta f=f_r-3f_1$
  in kHz, and the ratio $\sigma/\sigma_0$ of rms bunch length to the natural bunch length. }
   \label{fig:fig1}
\end{figure}

\begin{figure}[htb]
   \centering
   \begin{minipage} [b]{.49\linewidth}
   \includegraphics[width=0.8\linewidth]{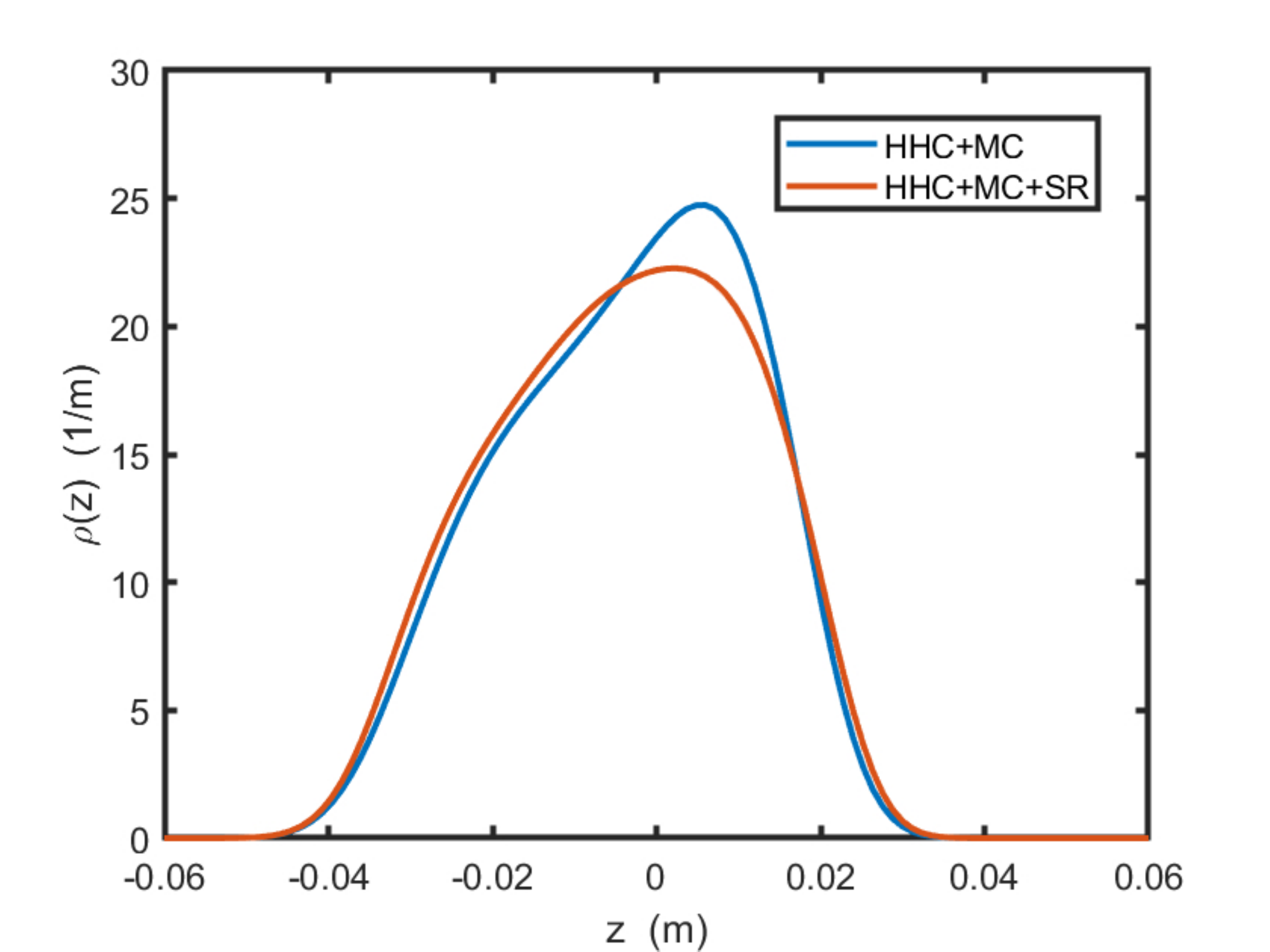}
  \caption{Complete fill, high $R/Q$, $\delta f=$\\ 317.8 kHz, $\sigma/\sigma_0 = 3.80$ (blue), 3.92 (red)}
   \label{fig:fig2}
   \end{minipage}
   \begin{minipage} [b]{.49\linewidth}
      \includegraphics[width=0.8\linewidth]{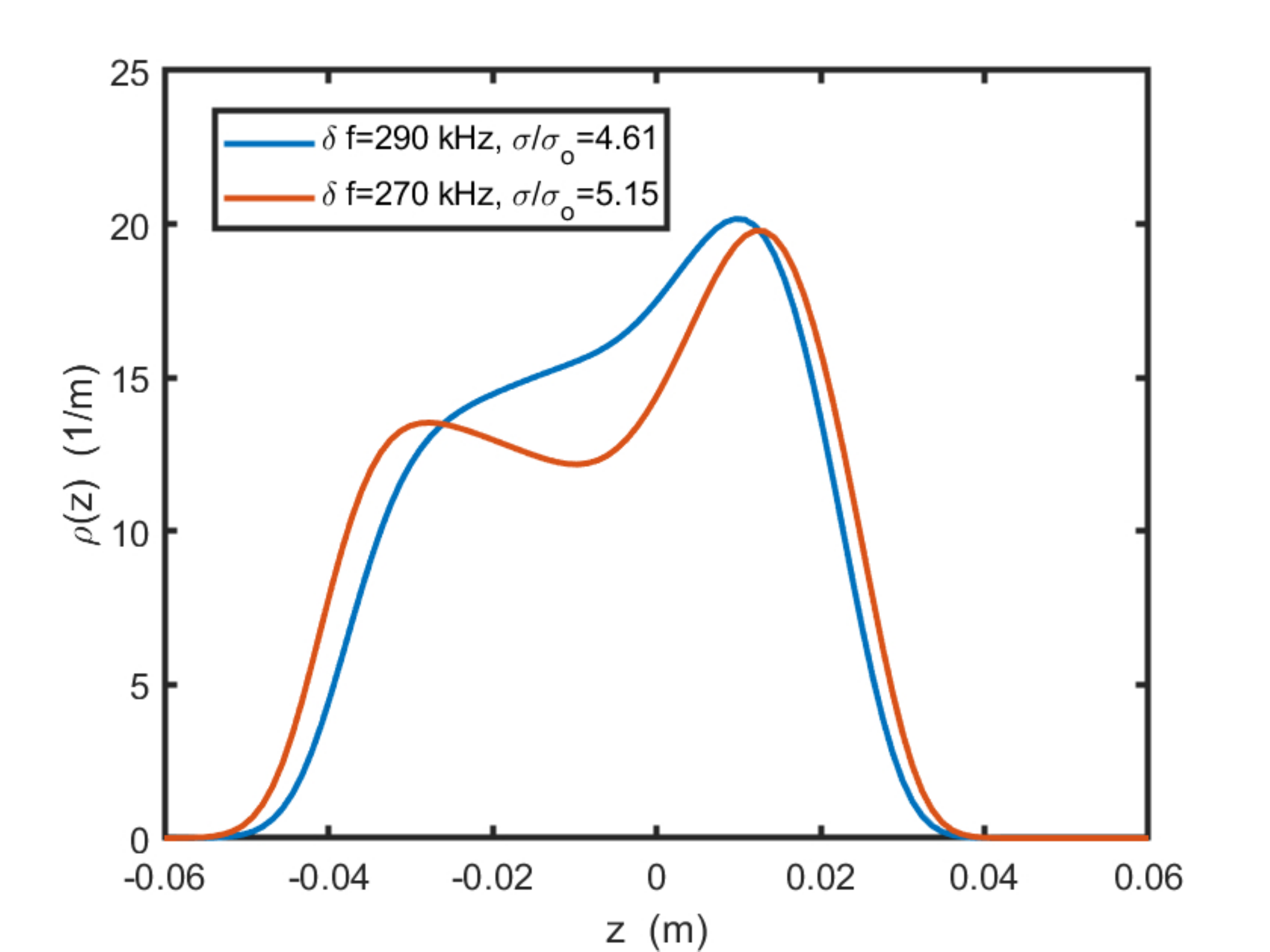}
  \caption{Complete fill, high $R/Q$,\\ with HHC+MC+SR. Two lower detunings.}
   \label{fig:fig3}
   \end{minipage}
\end{figure}

\subsection{The case of high $R/Q$ \label{subsection:high}}
\begin{figure}[htb]
   \centering
   \includegraphics[width=0.6\linewidth]{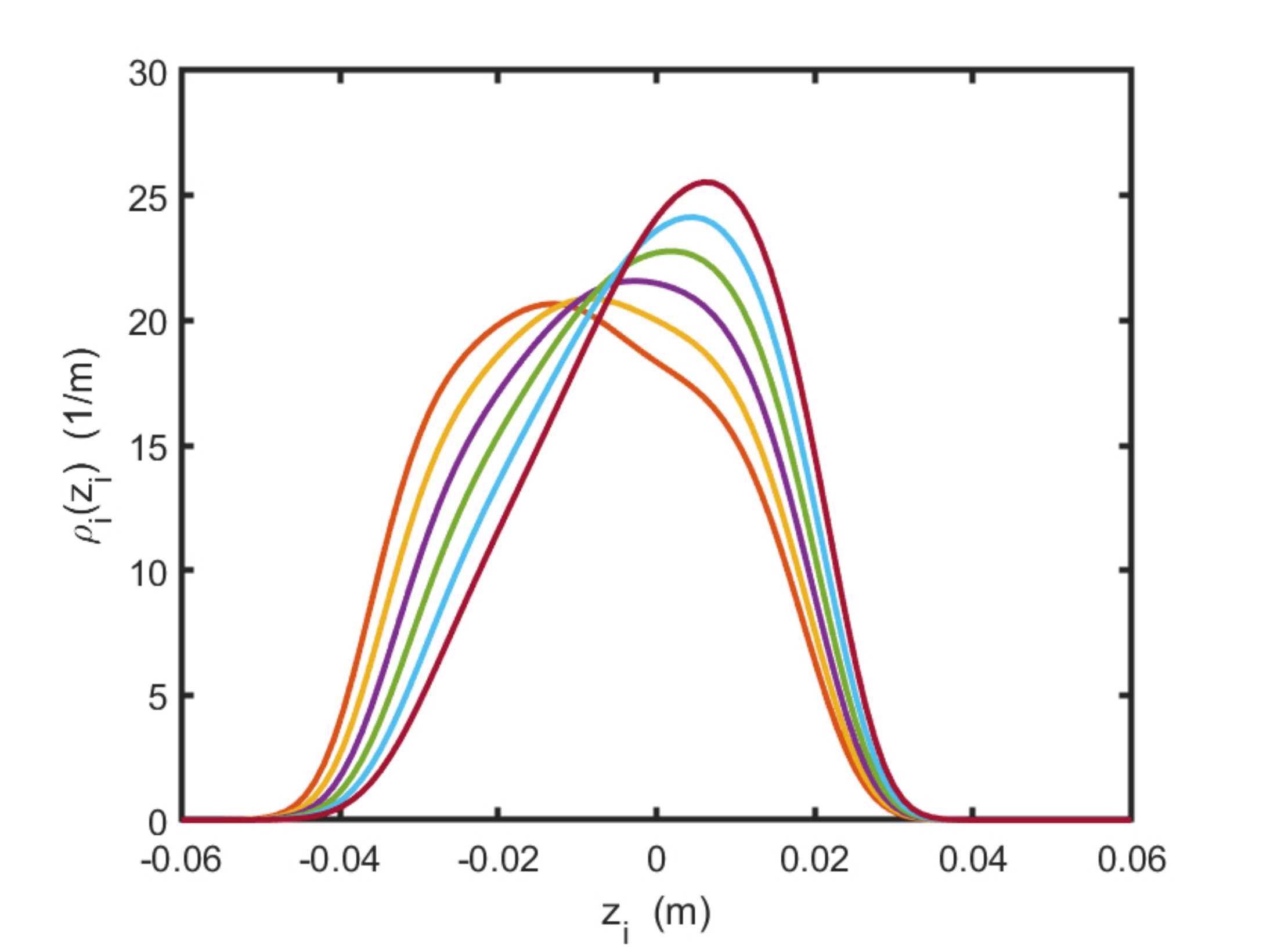}
  \caption{Fill with distributed gaps, high $R/Q$, HHC+MC+SR, $\delta f=317.8$ kHz.  }
   \label{fig:fig4}
\end{figure}
\begin{figure}[htb]
   \centering
   \begin{minipage} [b]{.49\linewidth}
   \includegraphics[width=\linewidth]{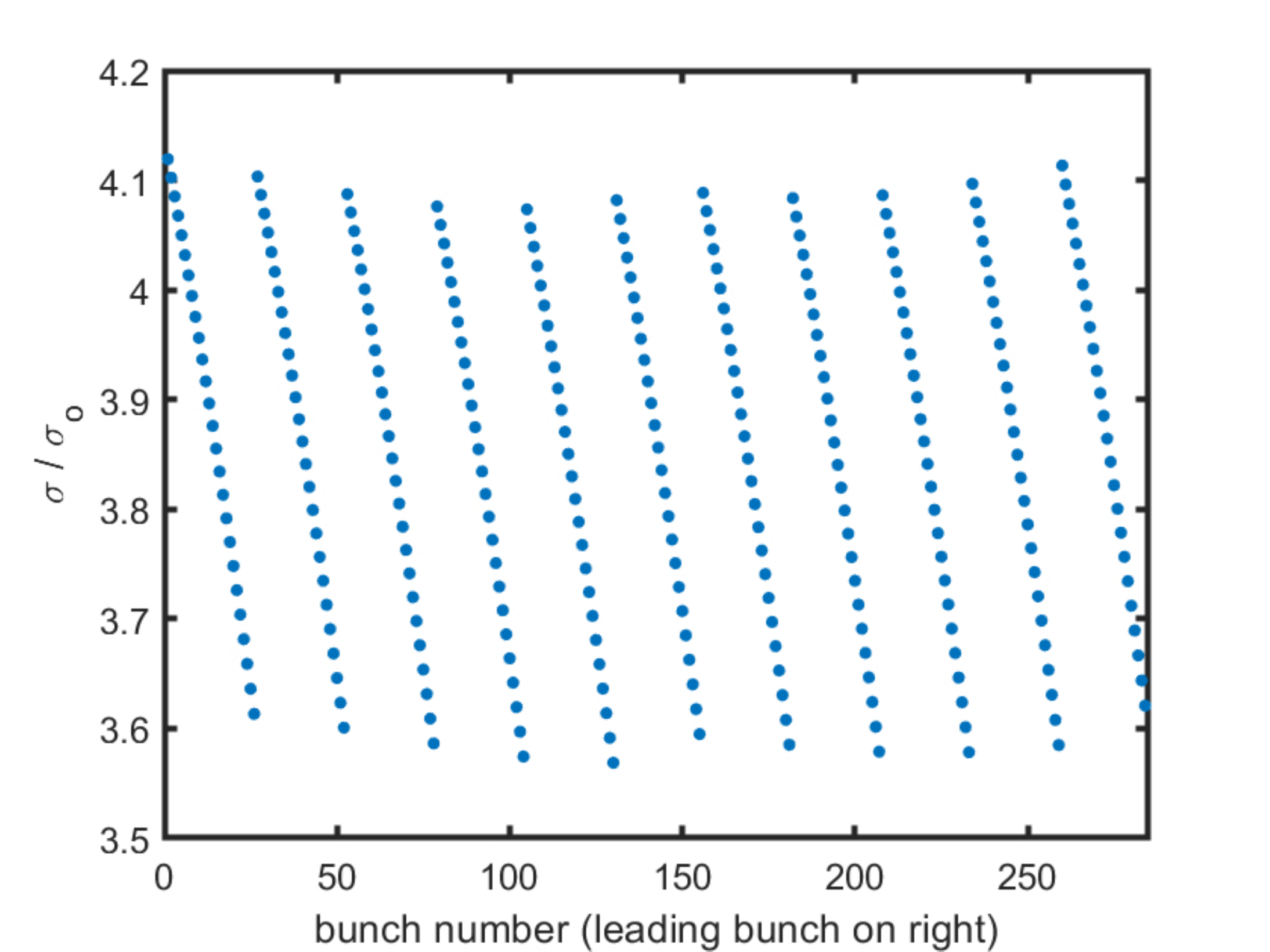}
  \caption{Case of Fig.\ref{fig:fig4}, ratio of bunch length\\ to natural length, vs. bunch number.}
   \label{fig:fig5}
   \end{minipage}
   \begin{minipage} [b]{.49\linewidth}
   \includegraphics[width=\linewidth]{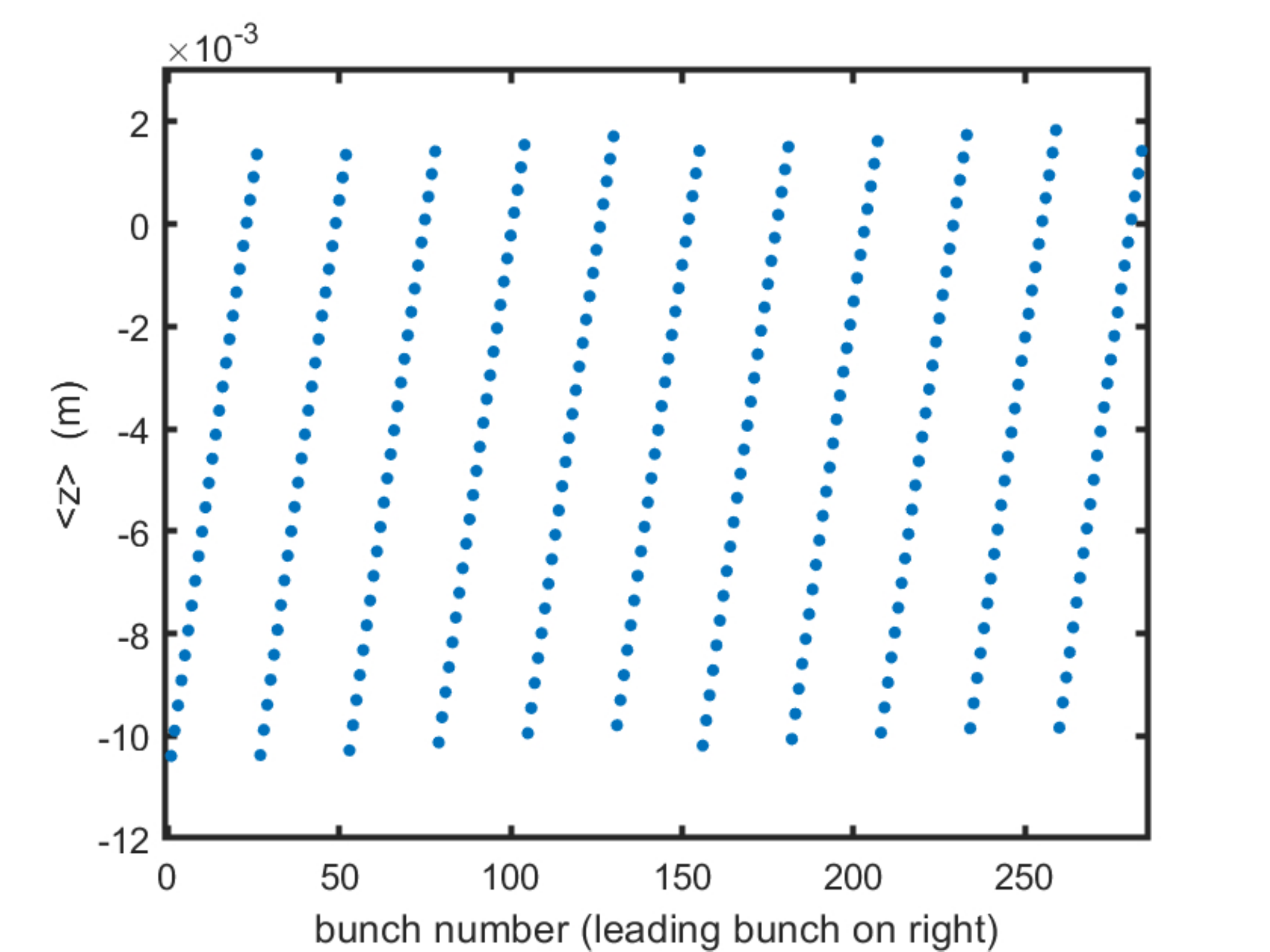}
   \caption{Case of Fig.\ref{fig:fig4}, bunch centroid \\ vs. bunch number.}
   \label{fig:fig6}
   \end{minipage}
\end{figure}
 \begin{figure}[htb]
   \centering
   \includegraphics[width=0.6\linewidth]{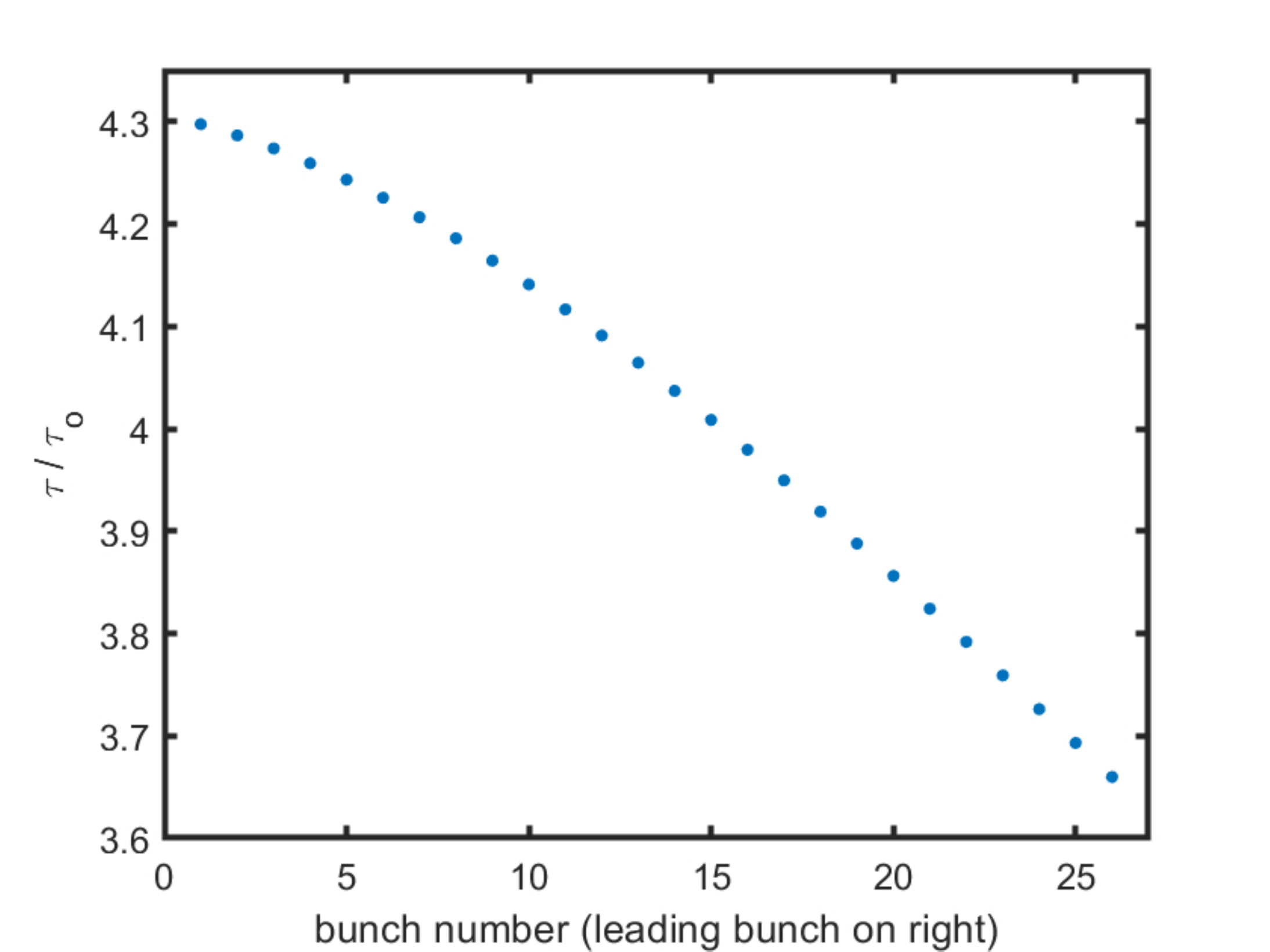}
  \caption{Case of Fig.\ref{fig:fig4}, increase of Touschek lifetime. }
   \label{fig:fig7}
   \end{figure}
The first step is to consider the complete fill, with all 328 buckets filled with the same  charge, and the entire wake coming from the HHC. Then every bunch comes out to have the same profile, even though that is not put in as a constraint. In Fig.\ref{fig:fig1} we show that profile for a decreasing sequence of detunings. The legend gives the detuning $\delta f$ in kHz and $\sigma/\sigma_0$, the ratio of the rms bunch length to the natural bunch length of 3.9 mm. This plot agrees with Fig.(3.255) in \cite{pdr}. With 201 mesh points per bunch, the CPU time is 6 seconds for each curve.

The next step, again for a complete fill, is to see the effect of the main cavity beam loading, which is to be compensated by adjustment of the generator voltage. As expected, the compensation is essentially perfect and the bunch profile is the same to graphical accuracy. It is given by the blue curve in Fig.\ref{fig:fig2} for the nominal detuning of 317.8 kHz from Table 1. Here the increase in bunch length is $\sigma/\sigma_0=3.79$

The compensation algorithm of \cite{prabII} did not converge with the desired energy loss of $U_0=315$ keV. Noticing that it did converge
in the work of \cite{prabI} which had a smaller value of $U_0$, we reduced the value and then increased it in steps: $U_0=260, 280, 300, 315$ keV. This procedure took 5 minutes.  All subsequent calculations were started with the result of a previous calculation and required  less than one minute of additional time each.

The compensation algorithm is the standard Gauss-Newton method for nonlinear least squares, although it was not recognized 
as such in \cite{prabII}. There are other  algorithms, such as the Levenberg-Marquardt method, which can be more robust concerning the starting guess.
Perhaps such a method could give a least squares solution directly for the desired $U_0$, but not necessarily in a shorter time.

Turning on the short range wake we get the red curve in Fig.\ref{fig:fig2}.  The short range force reduces the asymmetry of the bunch, and increases its rms length by 3\%. This is different from the effect of the short range wake in the broad band resonator model \cite{prabII},
and perhaps more reasonable.

Additional bunch lengthening through a decrease in detuning is a possibility for the machine, discussed in \cite{pdr}. The results of two smaller values are shown in Fig.\ref{fig:fig3}. The transition to overstretching, when two peaks appear, occurs between the two.

The partial fill anticipated for the machine, which has been called Fill C2 in \cite{pan}, has distributed gaps of 4 buckets each, with a total of 284 bunches.
There are 11 sub-trains, 9 with 27 bunches and 2 with 26, the latter two on opposite sides of the ring. All bunches have the same charge, chosen to give the desired average current of $500$ mA. Taking this case with the nominal detuning $\delta f=317.8$ kHz of Table 1, and
including HHC, MC, and SR,  we get the densities shown in Fig.\ref{fig:fig4}. There are 6 bunches in the plot out of a typical subtrain of 27 bunches. The one with maximum farthest to the right is nearest the front of the subtrain.

As was discovered in \cite{prabII}, the main cavity has a large influence when there are gaps in the train. The bunches near the front of
the sub-train resemble that of the complete fill, whereas those at the middle and back are broader and flatter. The distributions of bunch length increase and centroid displacement along the full train are shown in Fig.\ref{fig:fig5} and Fig.\ref{fig:fig6}. The bunch length increase and
the centroid displacement are both largest at the back of a sub-train.

The most interesting figure of merit is the increase in the Touscheck lifetime over the case without a harmonic cavity. This is plotted for a typical sub-train in Fig.\ref{fig:fig7}. The factor of increase, $\tau/\tau_0$, is not far from the length increase $\sigma/\sigma_0$. The strong variation along the sub-train should be an issue in determining the average beam lifetime, but that matter is beyond the scope of this work.

\begin{figure}[htb]
   \centering
   \includegraphics[width=0.6\linewidth]{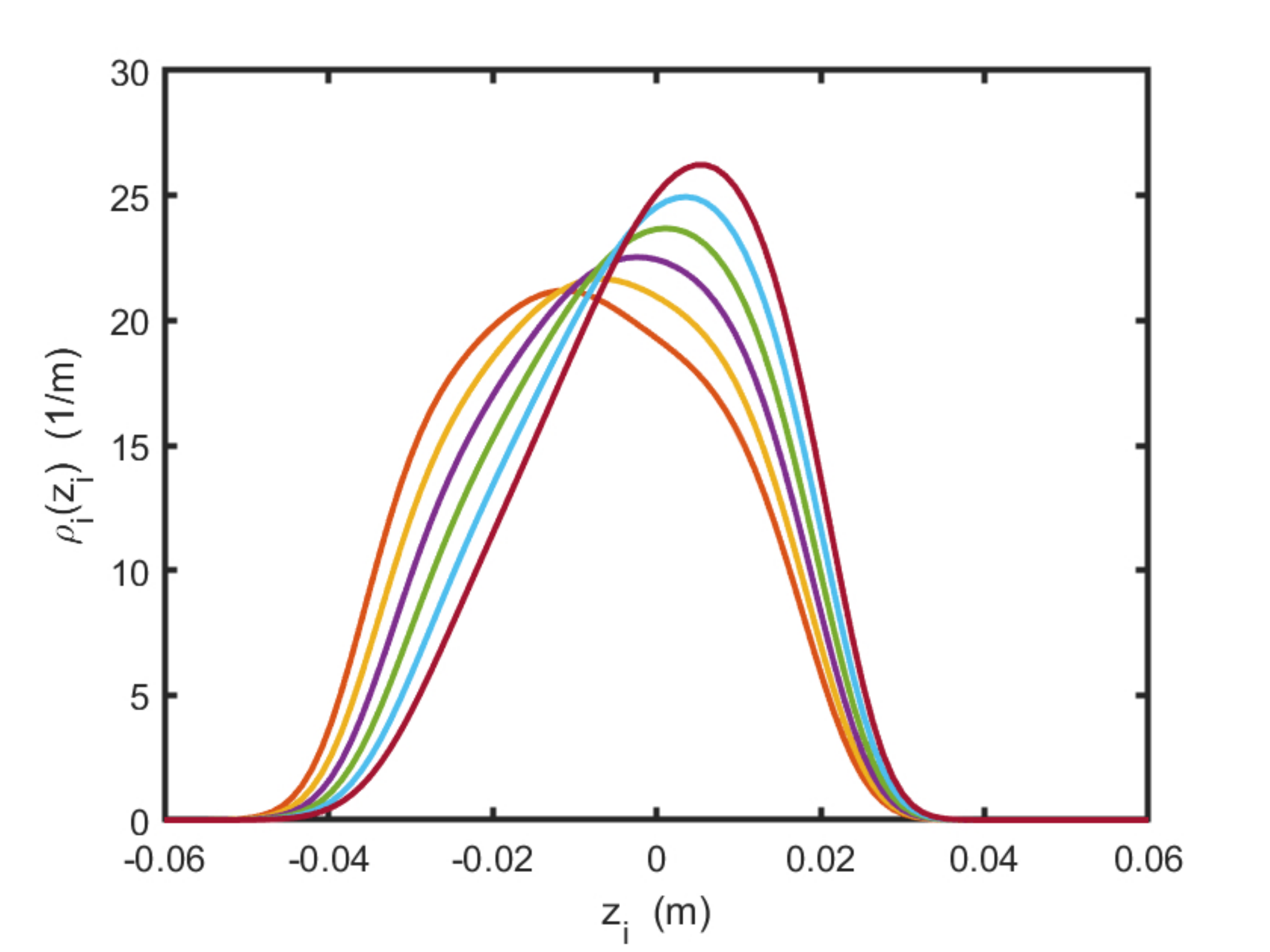}
  \caption{Fill with distributed gaps, HHC+MC+SR, $\delta f=330$ kHz.
  This result resembles Fig.(3.256) in \cite{pdr}.}
   \label{fig:fig8}
\end{figure}

Fig.\ref{fig:fig4} is to be compared with Fig.(3.256) in \cite{pdr}, generated from a  macro-particle simulation with rather severe noise. This plot is for a case different from ours in that it does not include the short range wake and most probably has a different account of the main cavity beam loading, which is mentioned but not described.
Also, the detuning is not specified exactly
but is said to be ``right below the onset of the overstretching instability".   Our Fig.\ref{fig:fig4} has more broadening, especially at the back of the train. For whatever reason, to get a result with a fair resemblance to Fig.(3.256) we have to {\it increase} the detuning to 330 kHz, getting the result in
Fig.\ref{fig:fig8}. Here the results for bunch lengthening, centroid position, and lifetime increase are consistent with the results in Fig.(3.256) of \cite{pdr} but a bit more favorable.

 \subsection{High $R/Q$ with overstretching\label{subsection:high_over}}
\begin{figure}[htb]
   \centering
   \begin{minipage} [b]{.49\linewidth}
   \includegraphics[width=\linewidth]{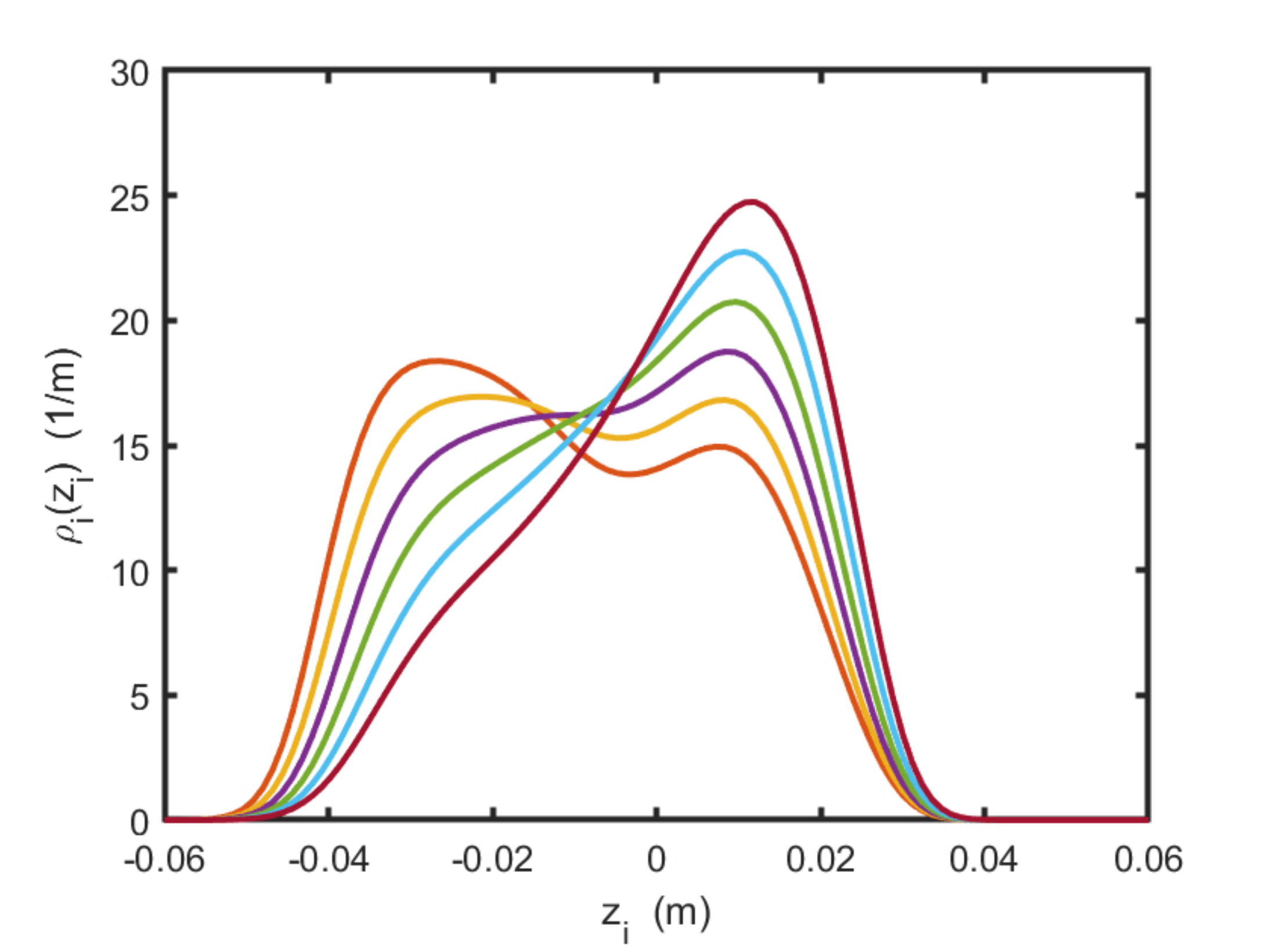}
  \caption{Fill with distributed gaps, \\ HHC+MC+SR, $\delta f=290$ kHz.}
   \label{fig:fig9}
   \end{minipage}
   \begin{minipage} [b]{.49\linewidth}
   \includegraphics[width=\linewidth]{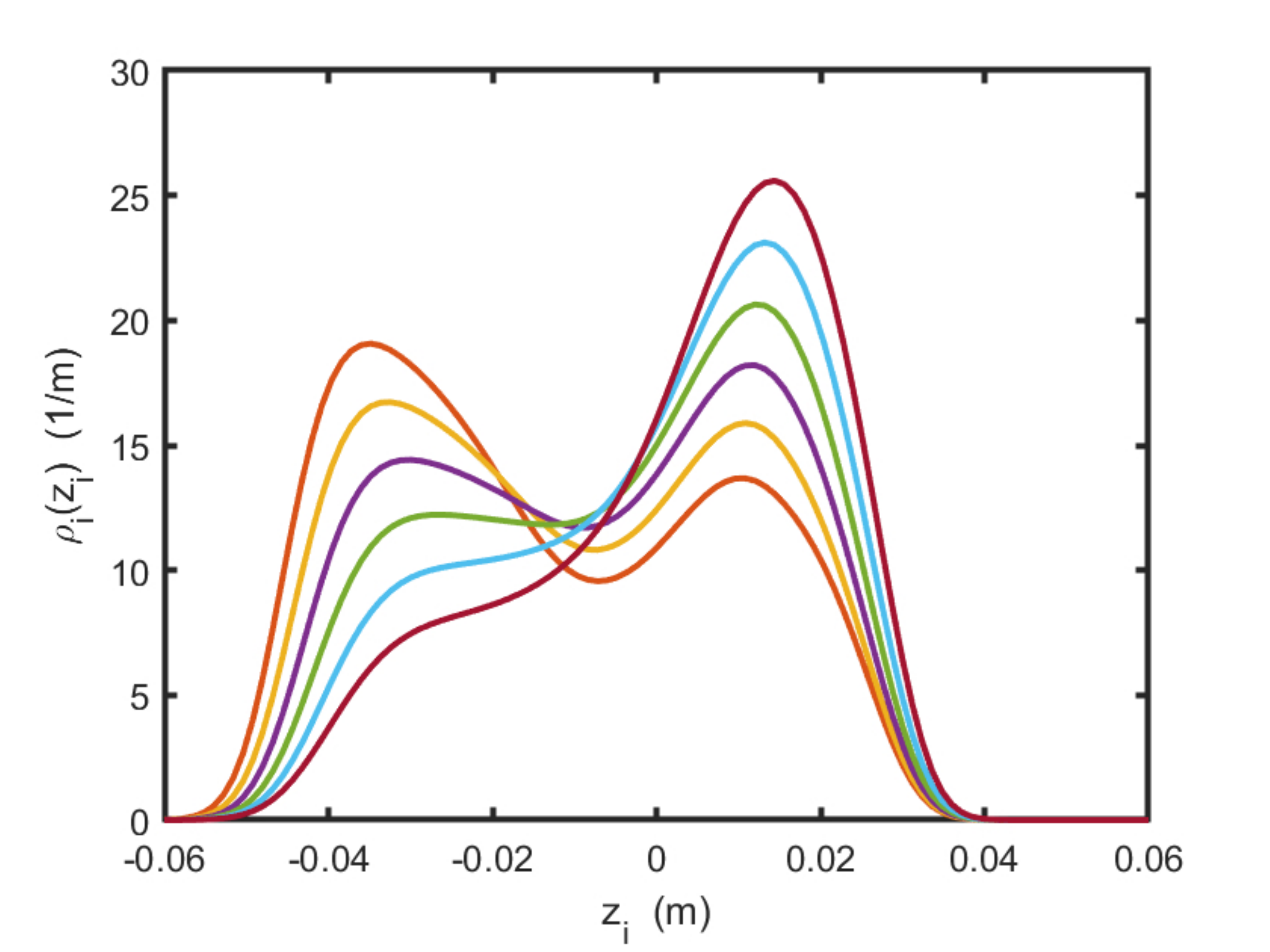}
   \caption{Fill with distributed gaps,\\ HHC+MC+SR, $\delta f=270$ kHz.}
   \label{fig:fig10}
   \end{minipage}
\end{figure}

\begin{figure}[htb]
   \centering
   \begin{minipage} [b]{.49\linewidth}
   \includegraphics[width=\linewidth]{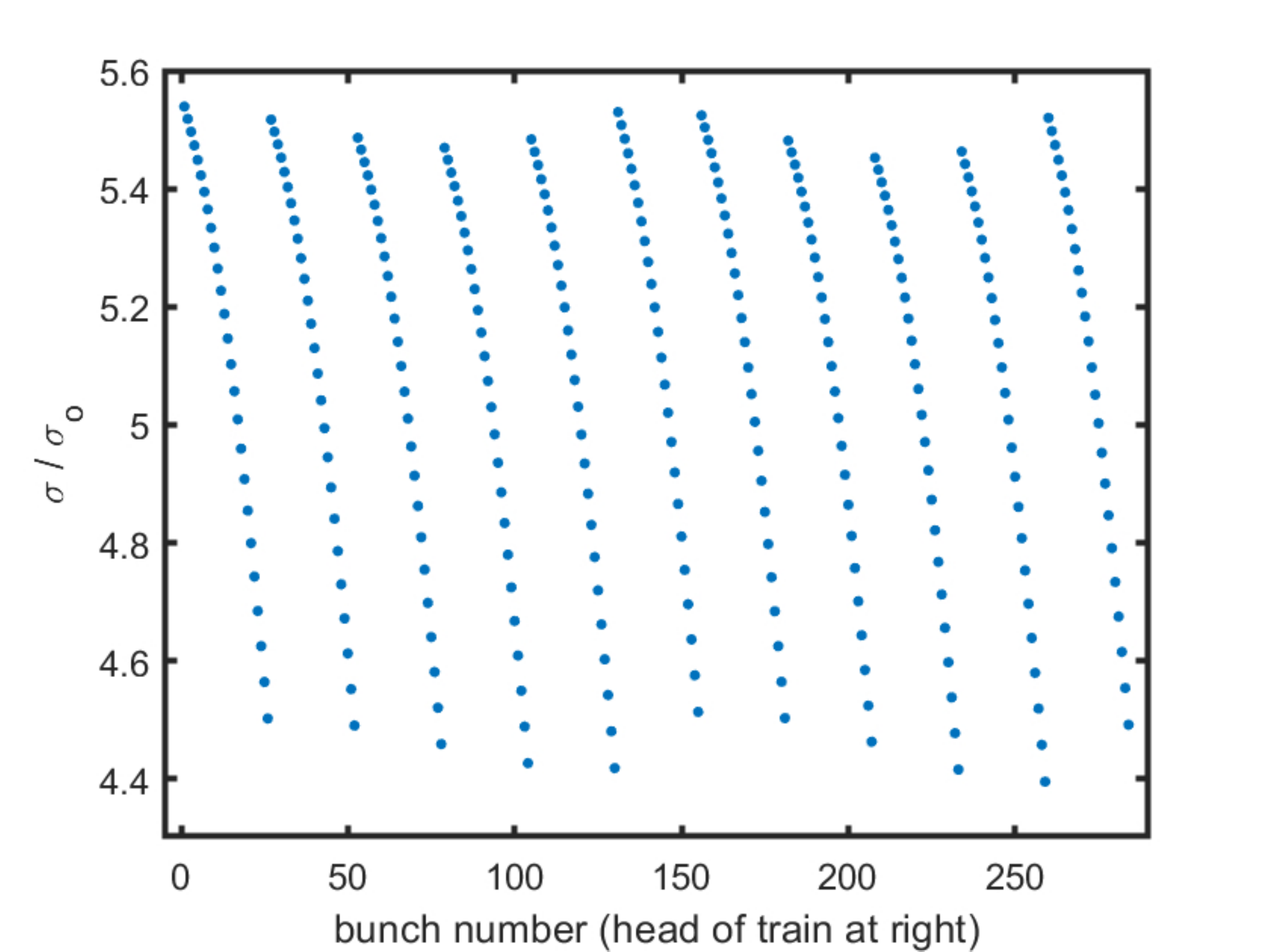}
  \caption{Fill with distributed gaps, \\ HHC+MC+SR, $\delta f=290$ kHz.}
   \label{fig:fig11}
   \end{minipage}
   \begin{minipage} [b]{.49\linewidth}
   \includegraphics[width=\linewidth]{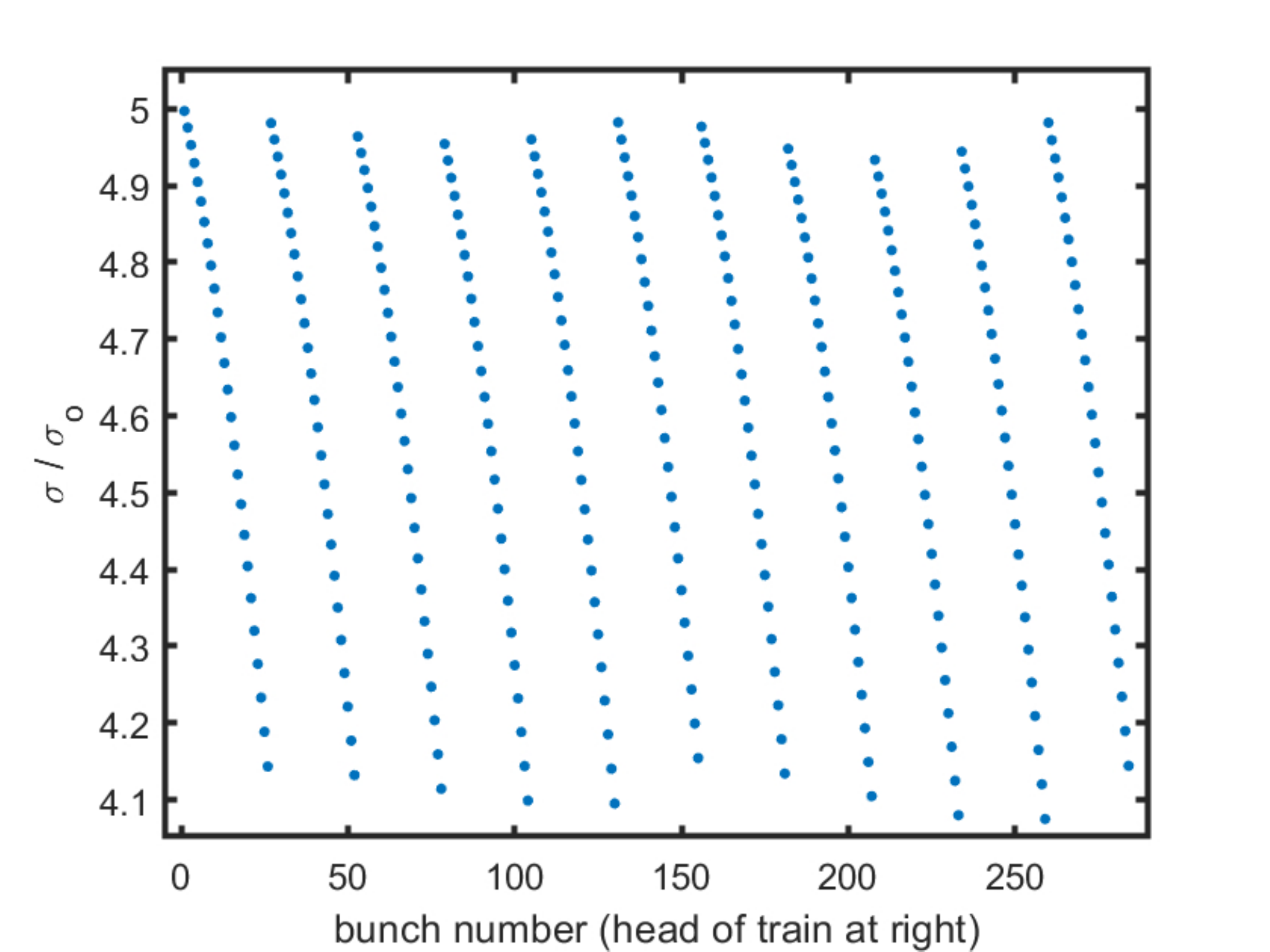}
   \caption{Fill with distributed gaps,\\ HHC+MC+SR, $\delta f=270$ kHz.}
   \label{fig:fig12}
   \end{minipage}
\end{figure}

 What is the effect of overstretching with the partial fill, as compared to the result of Fig.\ref{fig:fig4} for the complete fill?  It turns out that the threshold for overstretching is at larger $\delta f$ than Fig.\ref{fig:fig3} would indicate for bunches at the back of a sub-train,
 but similar to Fig.\ref{fig:fig3} for bunches at the front. Figs.\ref{fig:fig9} and \ref{fig:fig10} show the densities for the same detunings as in Fig.\ref{fig:fig4}. Figs.\ref{fig:fig11} and \ref{fig:fig12} show the corresponding bunch length increases.

 This shows that there is no profit in overstretching beyond a certain point. Fig.\ref{fig:fig12} displays more undesirable bunch distortion
 than \ref{fig:fig11} without much increase in the average bunch length.

 It took only 40 seconds to produce Fig.\ref{fig:fig9} starting with the solution of Fig.\ref{fig:fig4}, and another 40 seconds to make
 Fig.\ref{fig:fig10} starting with  Fig.\ref{fig:fig9}. Through Anderson acceleration we have gained a remarkable advance in technique compared to Ref.\cite{prabII} in which these solutions  could not be produced at all.

\subsection{The case of low $R/Q$ \label{subsection:low}}
\begin{figure}[htb]
   \centering
   \includegraphics[width=0.6\linewidth]{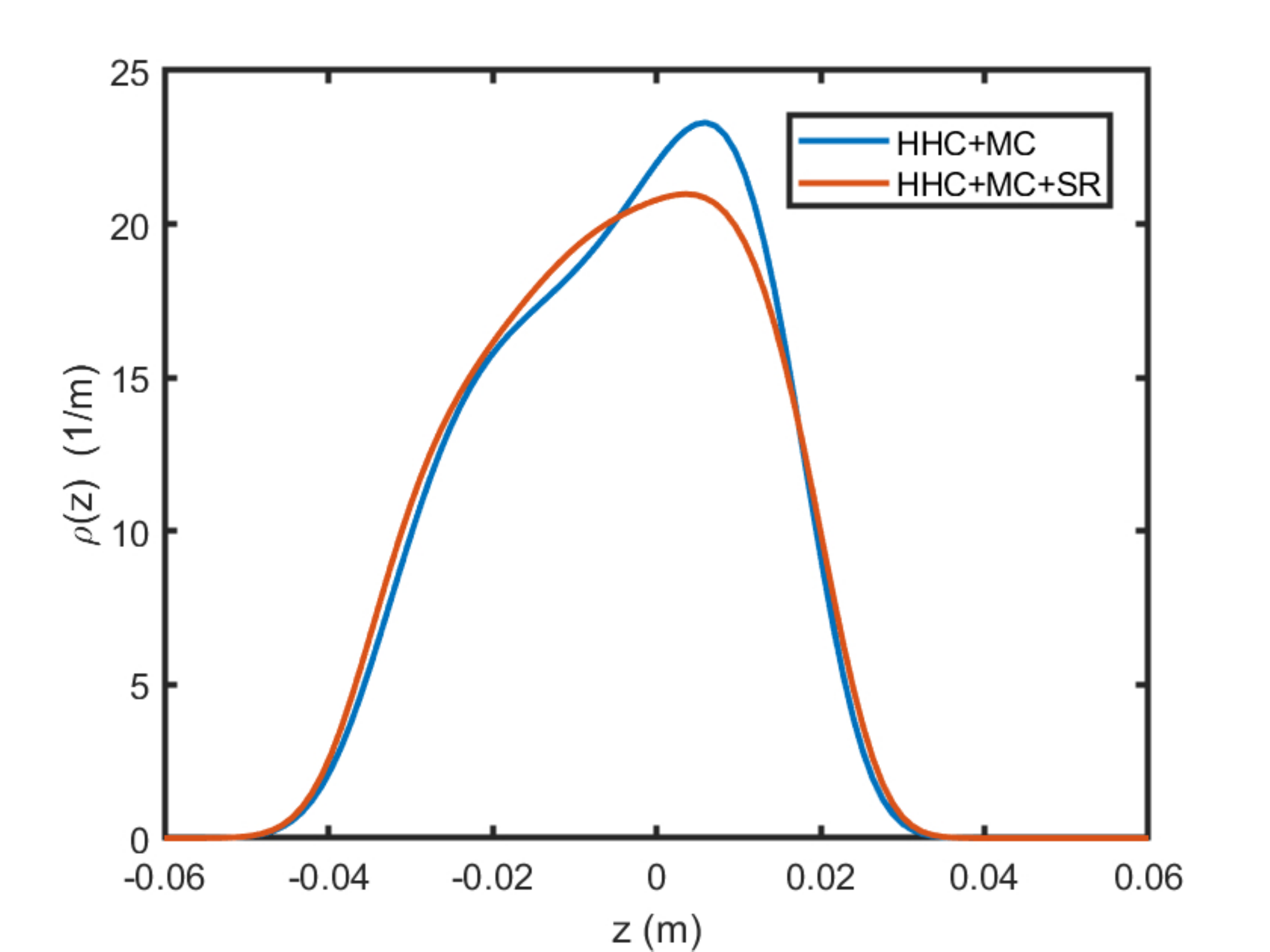}
  \caption{Complete fill, low $R/Q$, HHC+MC in the blue curve, HHC+MC+SR in the red curve. Detuning 164.7 kHz from Table 1. $\sigma/\sigma_0 = 3.96$ (blue), 4.07 (red)}
   \label{fig:fig13}
\end{figure}
\begin{figure}[htb]
   \centering
   \includegraphics[width=0.6\linewidth]{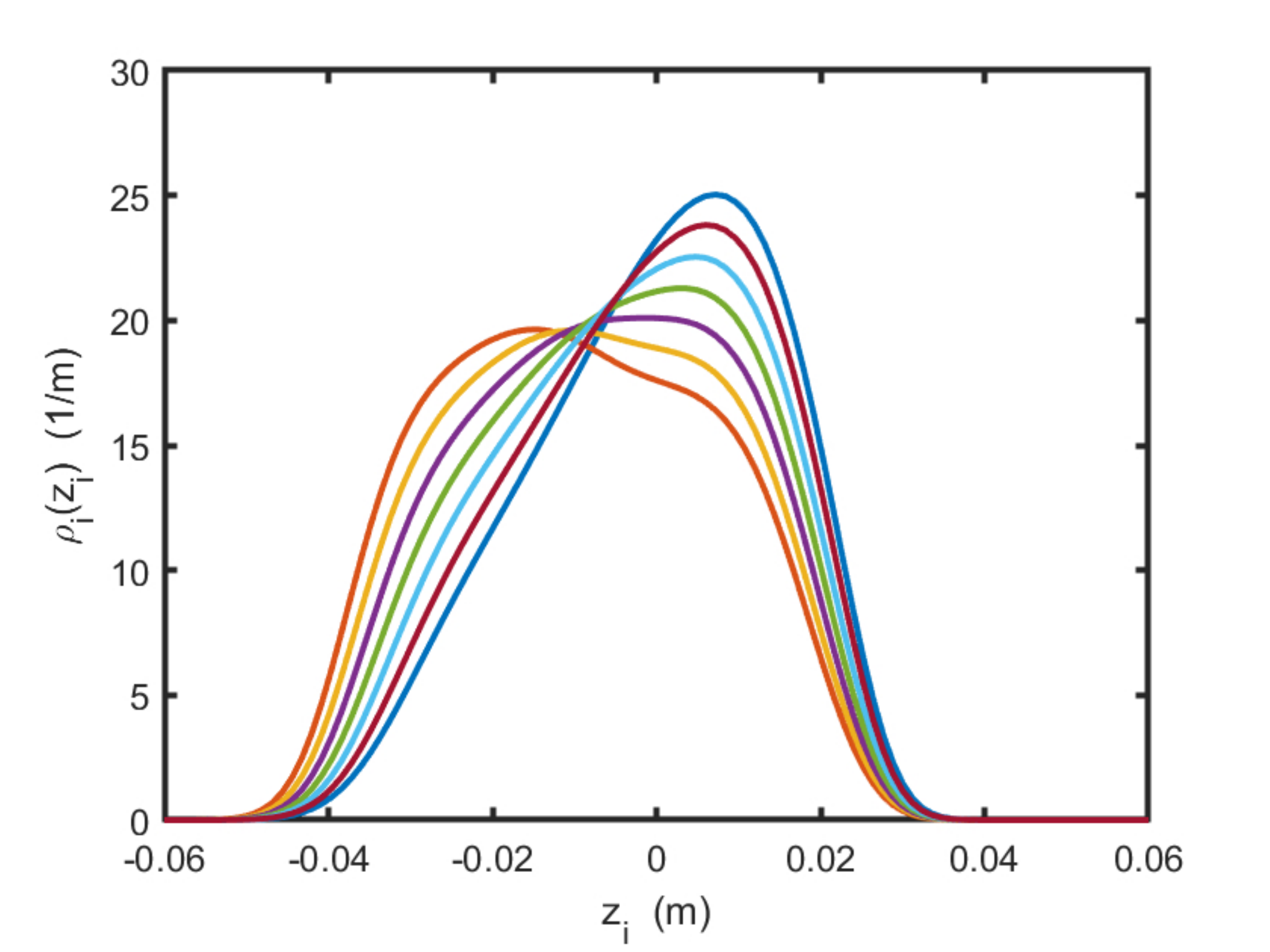}
  \caption{Fill with distributed gaps, low $R/Q$, HHC+MC+SR, $\delta f=164.7$ kHz.  }
   \label{fig:fig14}
\end{figure}

For low $R/Q$ the results for a complete fill are given in Fig.\ref{fig:fig13}.
Passing from this solution to the case of a partial fill we find the pattern of Fig.\ref{fig:fig14}.

The preliminary design report \cite{pdr} expresses an interest in running
this case with 20\% overstretching, which is illustrated with a macro-particle simulation in Fig.(3.257). We obtain
the closely similar result of Fig.\ref{fig:fig14} with a detuning of 140 kHz, reduced from 164.7 kHz.
\begin{figure}[htb]
   \centering
   \includegraphics[width=0.6\linewidth]{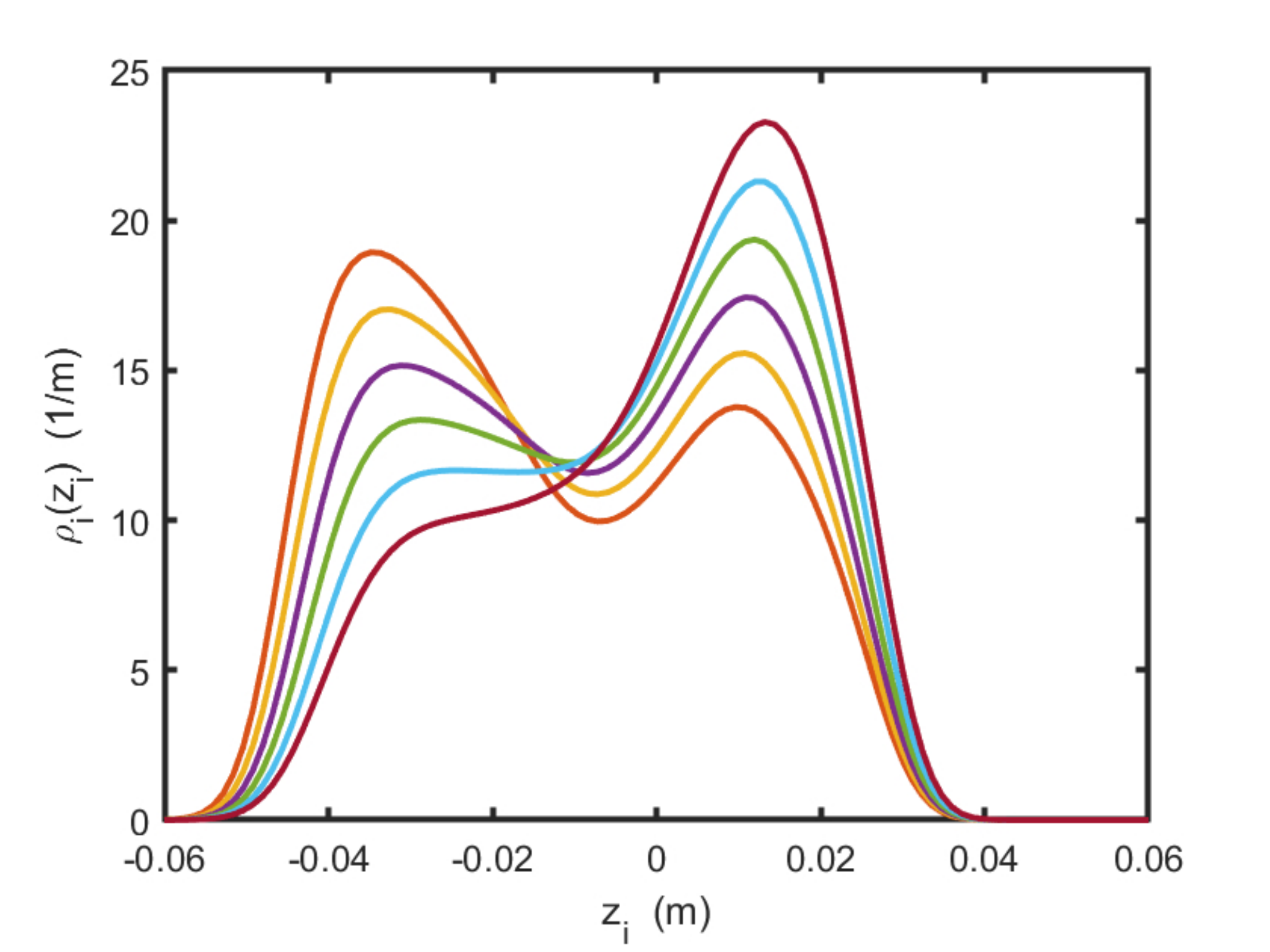}
  \caption{Fill with distributed gaps, low $R/Q$, HHC+MC+SR, $\delta f=140$ kHz.  }
   \label{fig:fig15}
\end{figure}
The corresponding outcomes for bunch lengthening, centroid distribution, and Touschek lifetime increase are plotted in Fig.\ref{fig:fig16}, Fig.\ref{fig:fig17}, and Fig.\ref{fig:fig18}.
\begin{figure}[htbp]
   \centering
   \begin{minipage} [b]{.49\linewidth}
   \includegraphics[width=\linewidth]{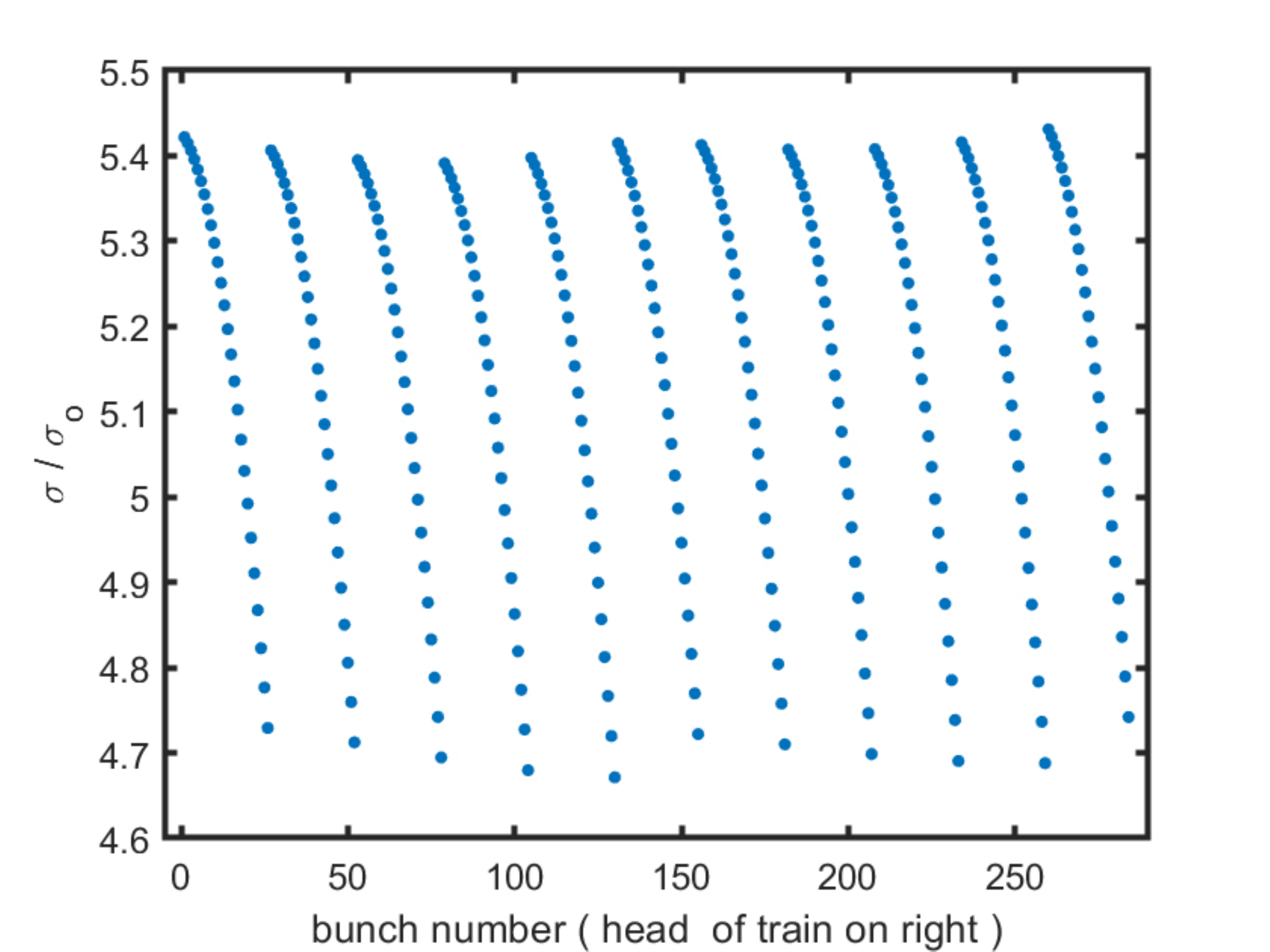}
  \caption{Case of Fig.\ref{fig:fig14}, ratio of bunch length\\ to natural length, vs. bunch number.}
   \label{fig:fig16}
   \end{minipage}
   \begin{minipage} [b]{.49\linewidth}
   \includegraphics[width=\linewidth]{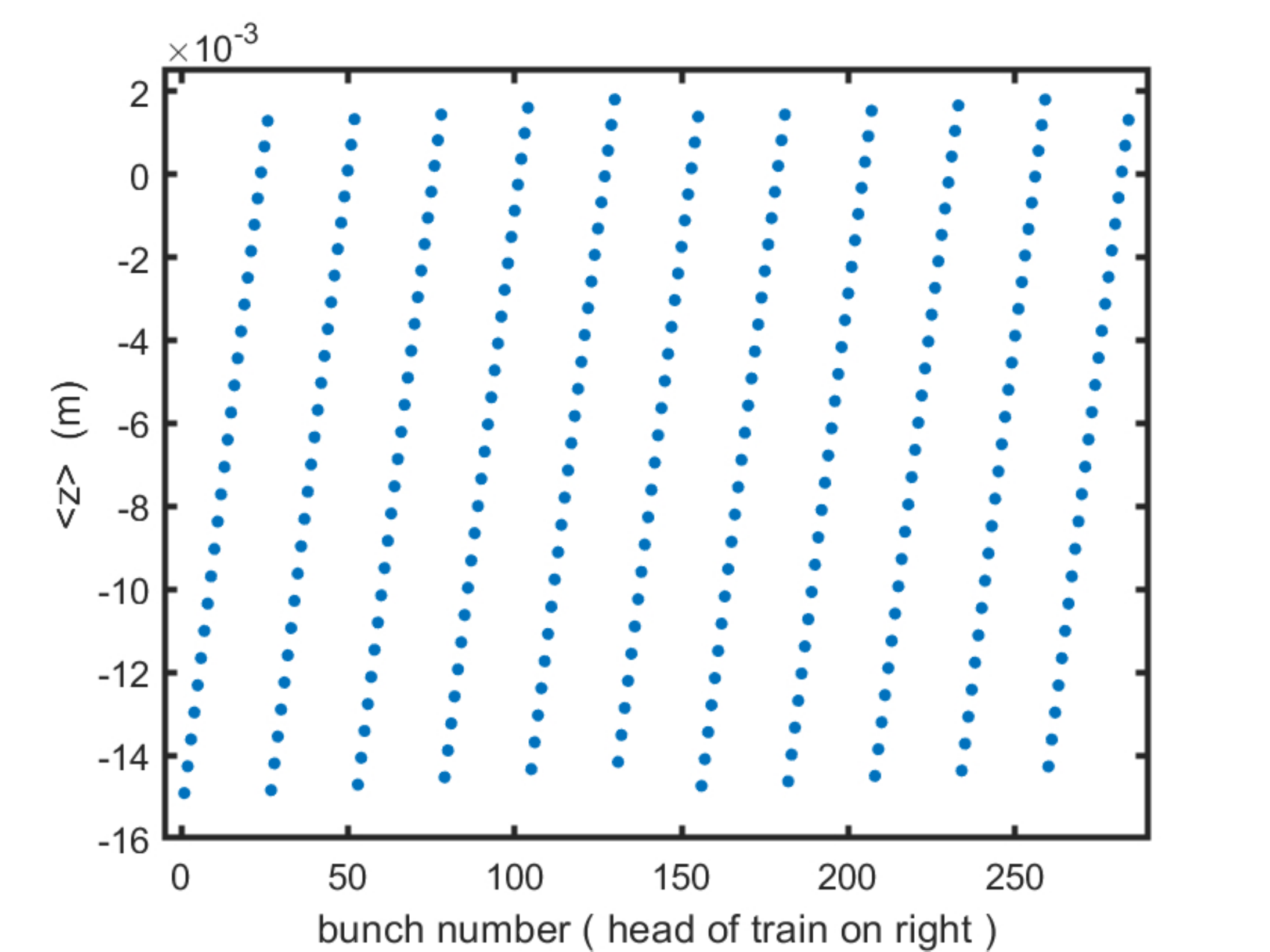}
   \caption{Case of Fig.\ref{fig:fig14}, bunch centroid \\ vs. bunch number.}
   \label{fig:fig17}
   \end{minipage}
\end{figure}
\begin{figure}[htbp]
   \centering
   \includegraphics[width=0.6\linewidth]{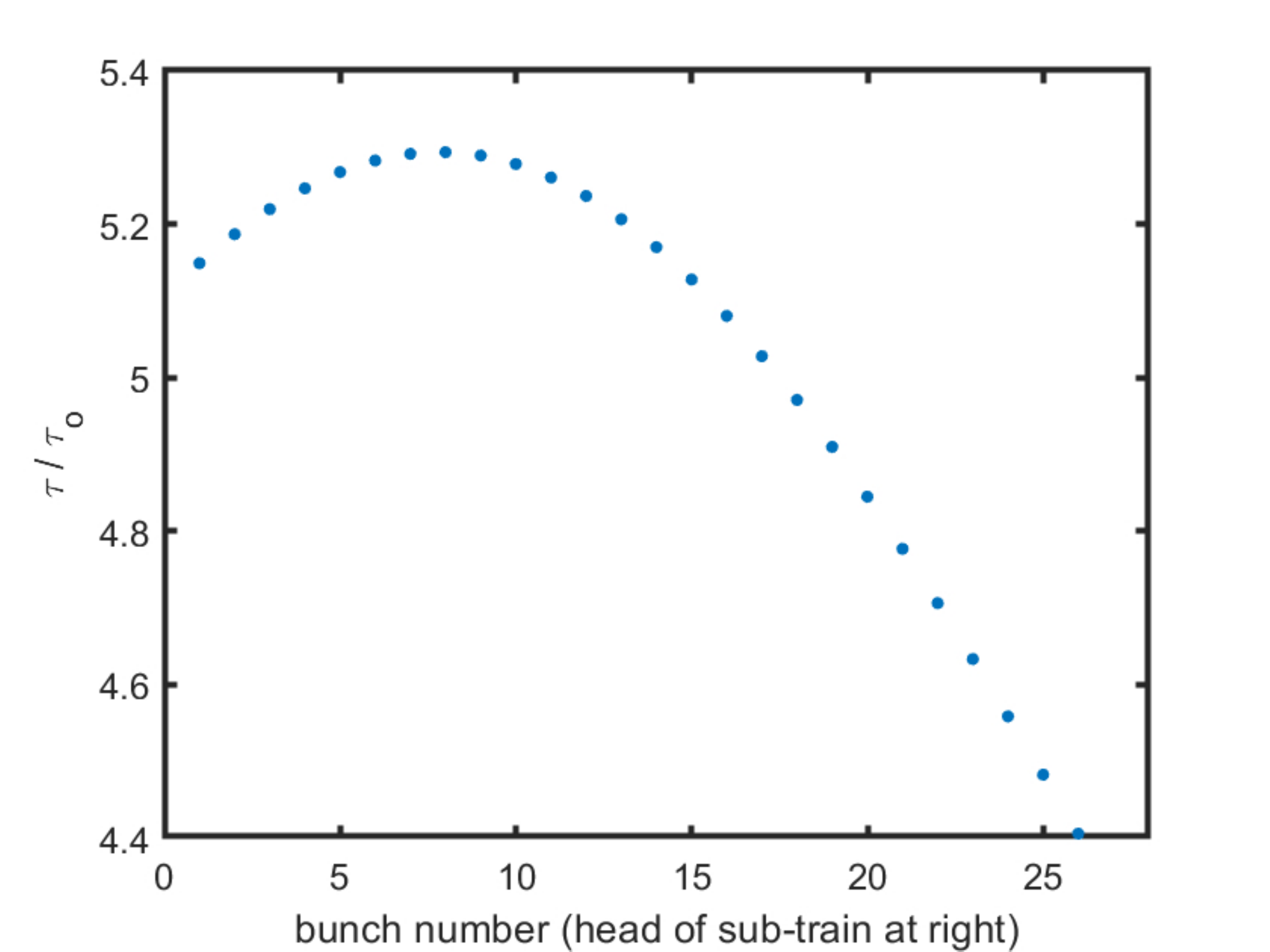}
  \caption{Case of Fig.\ref{fig:fig14}, increase of Touschek lifetime. }
   \label{fig:fig18}
   \end{figure}

\section{The relation of Anderson acceleration to Broyden's method \label{section:relation}}
By an argument of Eyert \cite{eyert}, revisited by Fang and Saad \cite{fang-saad}, Anderson's method is equivalent to a generalized form of Broyden's second method for updating
an approximation to the inverse Jacobian as in (\ref{broyden}). The usual Broyden method  imposes the secant condition,
\be
G_{k+1}\Delta f_k=\Delta x_k\ ,  \label{secant}
\ee
which is motivated by the linear approximation to $f$. A second condition requires that the Frobenius squared norm $\| G_{k+1}-G_k\|_F^2$ be minimum with respect to $G_{k+1}$, subject to condition (\ref{secant}). These two conditions lead uniquely to (\ref{broyden}).

A generalized Broyden method imposes  multi-secant conditions,
\be
G_k\Delta f_p=\Delta x_p\ ,\qquad p=k-m_k,\cdots,k-1\ ,    \label{multisecant}
\ee
thus taking account of previous iterates in the spirit of Anderson. Again imposing a minimum change of $G_k$, and taking $G_0=-I$, the updates $x_{k+1}=x_k-G_kf_k$ are found to coincide with Anderson's. The matrix $\mathcal {F}_k $ having the vectors $\Delta f_p$ as columns is assumed to have full rank.

\section{Conclusions and outlook \label{section:fine}}
I have described a new method to compute the equilibrium charge densities of an arbitrary bunch train with gaps, under the influence of an harmonic cavity (HHC), the main accelerating cavity (MC), and a realistic short range wake field (SR). It succeeds under more difficult conditions than previous methods, in fact for all conditions that arise in examples studied to date. Realized by a serial code on a laptop, the method takes only a few minutes for a thorough survey of the parameter space.

As an example, the parameter set for ALS-U in the Preliminary Design Report was adopted.  Results similar to those obtained by macro-particle simulations in the report could be obtained with a reasonable choice of detuning parameters. The report does not specify detuning exactly, and
the physical model is different in not including the short range wake, and may have a different treatment of the MC beam loading.

This paper introduces the Anderson iterative method that is probably new to accelerator physics, and which seems very promising for further applications
in the field. It is especially interesting for problems falling under rubrics such as ``self-consistency" or ``phase space matching", often formulated
in terms of nonlinear integral or differential equations.

There are  large-scale applications of Anerson's method and related ideas in the literature of {\it ab initio} quantum mechanical
calculations of material and molecular properties \cite{dederichs, eyert, fang-saad, kresse, pulay}. These are based on the Kohn-Sham density functional formalism \cite{kohn}, which is similar in spirit
if not in specifics to systems arising from the nonlinear Vlasov equation. It may be profitable to keep an eye on this work to see if there are
any lessons to be learned for accelerator physics. Also, ongoing efforts by numerical analysts are interesting, specifically for Anderson acceleration \cite{desterck}.
\section{Acknowledgements \label{acknow}}
I learned of Anderson acceleration through an inspiring colloquium  for the University of New Mexico by Prof. Hans De Sterck, University of Waterloo. I thank Dr. Dan Wang for
a copy of her wake potential for ALS-U. This work was supported in part by the U. S. Department of Energy, Contract DE-AC03-76SF00515.


\begin{thebibliography} {99}
\bibitem{prabI} R. Warnock and M. Venturini, Equilibrium of an arbitrary bunch train in presence of a passive harmonic
cavity: Solution through coupled Ha\" issinski equations, Phys. Rev. Accel. Beams {23}, 064403 (2020).
\bibitem{prabII} R. Warnock, Equilibrium of an arbitrary bunch train in the presence of multiple resonator wake fields,
Phys. Rev. Accel. Beams {\bf 24}, 024401 (2021).
\bibitem{bobkarl} R. Warnock and K. Bane, Numerical solution of the Ha\"issinski equation for the equilibrium state of a stored
electron beam, Phys. Rev. Accel. Beams {\bf 21} 124401 (2018).
\bibitem{hefei} T. He, W. Li, Z. Bai, L. Wang, Numerical solution of the coupled Ha\"issinski equations for the equilibrium state of an arbitrary
bunch train in an electron storage ring, Nucl. Inst. Meth. Phys. Res. A {\bf 1006}, 165434 (2021); Longitudinal equilibrium density distribution
of arbitray filled bunches in presence of a passive harmonic cavity and the short range wakefield, Phys. Rev. Accel. Beams {\bf 24}, 044401 (2021).
\bibitem{dederichs} P. H. Dederichs and R. Zeller, Self-consistency iterations in electronic-structure calculations, Phys. Rev B {\bf 28}, 5462 (1983).
\bibitem{ortega} J. M. Ortega and W. C. Rheinboldt, ``Iterative Solution of Nonlinear Equations
in Seversl Variables", (Academic Press, New York, 1970).
\bibitem{boggs} P. T. Boggs, The solution of nonlinear systems of equations by A-stable integration techniques,
SIAM J. Numer. Anal. {\bf 8}, 767 (1971).
\bibitem{broyden} C. G. Broyden, A class of methods for solving nonlinear simultaneous equations, Math. Comp. {\bf 19}, 577 (1965).
\bibitem{kelley} C. T. Kelley, ``Iterative Methods for Linear and Nonlinear Equations", (SIAM, Philadelphia, 1995).
\bibitem{brezinski} C. Brezinski, Convergence acceleration during the 20th century, J. Comput. Appl. Math. {\bf 122}, 1 (2000).
\bibitem{anderson} D. G. Anderson, Iterative procedures for nonlinear integral equations, J. Assoc. Comput. Mach., {\bf 12}, 547 (1965).
\bibitem{walker-ni} H. F. Walker and P. Ni, Anderson acceleration for fixed-point iteration, SIAM J. Numer. Anal. {\bf 40}, 1715 (2011).
\bibitem{pan} Z. Pan, S. De Santis, T. Hellert, C. Steier, C. Sun, C. Tang, and M. Venturini, Beam-loading transients and bunch shape
in the operation of passive harmonic cavities in the ALS-U, Proc. IPAC2018, Vancouver BC, Canada.
\bibitem{eyert} V. Eyert, A comparitive study on methods for convergence acceleration of iterative vector sequnces, J. Comp. Phys. {\bf 124}, 271 (1996).
\bibitem{fang-saad} H. Fang and Y. Saad, Two classes of multisecant methods for nonlinear acceleration, Numer. Linear Algebra Appl. {\bf 16}, 197 (2009).
\bibitem{pdr} https://drive.google.com/file/d/1B\_jMmOJkZYFLZ3PnCxYOietSPYAtwV03/view; see pp. 223-273.
\bibitem{dwang} Dan Wang, Lawrence Berkeley National Laboratory, private communication. This is from
work in progress, and may not be the final result.
\bibitem{kresse} G. Kresse and J. Furthm\" uller, Effective iterative schemes for {\it ab initio} total-energy calculations
using a plane-wave basis set, Phys. Rev. B {\bf 54}, 11169 (1996).
\bibitem{pulay} P. Pulay, Convergence acceleration of iterative sequnces. The case of SCF iteration., Chem. Phys. Lett. {\bf 73}, 393 (1980).
\bibitem{kohn} W. Kohn and L. Sham, Phys. Rev. {\bf 140} A1133 (1965).
\bibitem{desterck} H. De Sterck and Y.He, On the asymptotic linear convergence speed of Anderson acceleration, Nestorov
acceleration, and nonlinear GEMRES, SIAM J. Sci. Comput. (2020).
\end{thebibliography}
\end{document}